\documentclass[preprint]{aastex}
\begin{document}
\title{Discovery of 9 Ly$\alpha$ emitters at
       redshift $z\sim3.1$ using narrow-band imaging and VLT
       spectroscopy\footnote{Based on observations collected at the
       European Southern Observatory, Cerro Paranal, Chile; ESO
       programmes 63.N-0530 and 63.I-0007}}
\author{R.-P. Kudritzki\altaffilmark{2,3} and R. H. M\'endez}
\affil{Munich University Observatory, Scheinerstr. 1, 
       81679 Munich, Germany}
\email{kudritzki@usm.uni-muenchen.de, mendez@usm.uni-muenchen.de}
\author{J. J. Feldmeier and R. Ciardullo}
\affil{Dept. of Astron. and Astrophys., Penn State
       Univ., 525 Davey Lab, Univ. Park, PA 16802}
\author{G. H. Jacoby}
\affil{Kitt Peak National Observatory, P.O.Box 26732, Tucson, 
       AZ 85726}
\author{K. C. Freeman}
\affil{Mt. Stromlo and Siding Spring Observatories, 
       Weston Creek P.O., ACT 2611, Australia}
\author{M. Arnaboldi and M. Capaccioli}
\affil{Osservatorio Astronomico di Capodimonte, 
       V. Moiariello 16, Napoli 80131, Italy}
\author{O. Gerhard}
\affil{Astronomisches Institut, Universit\"at Basel, CH-4102
       Binningen, Switzerland}
\and
\author{H. C. Ford}
\affil{Phys. and Astron. Dept., John Hopkins Univ., 
                      Homewood Campus, Baltimore, MD 21218}
\altaffiltext{2}{Max-Planck-Institut f\"ur Astrophysik,
                 Karl-Schwarzschild-Str. 1, 85740 Garching, Germany}
\altaffiltext{3}{Steward Observatory, Univ. of Arizona,
                 933 N. Cherry Av., Tucson, AZ 85721}

\newpage

\begin{abstract}
Narrow-band imaging surveys aimed at detecting the faint emission from
the 5007 \AA \ [O III] line of intracluster planetary nebulae in Virgo
also probe high redshift $z\sim3.1$ Ly$\alpha$ emitters.  Here we report 
on the spectroscopic identification of 9 Ly$\alpha$ emitters at
$z=3.13$ with fluxes between $2 \times 10^{-17}$ and $2 \times 10^{-16}$
erg cm$^{-2}$ s$^{-1}$ obtained with the FORS spectrograph at Unit 1
of the ESO Very Large Telescope (VLT UT1).

The spectra of these high redshift objects show a narrow, isolated 
Ly$\alpha$ emission with very faint (frequently undetected) continuum, 
indicating a large equivalent width. No other features
are visible in our spectra. Our Ly$\alpha$ emitters are quite 
similar to those found by Hu (1998), Cowie \& Hu (1998) 
and Hu et al. (1998). For a flat universe with 
H$_0$=70 km s$^{-1}$ Mpc$^{-1}$ and q$_0$=0.5 ($\Omega_\Lambda$=0),
the Ly$\alpha$ luminosity of the brightest source is
$1.7 \times 10^9$ L$\odot$, and the comoving space density of 
the Ly$\alpha$ emitters in the searched volume is 
$5 \times 10^{-3}$ Mpc$^{-3}$. 

Using simple population synthesis 
models, on the assumption that these sources are regions of 
star formation, we conclude that the nebulae are nearly
optically thick and must have a very low dust content, in order 
to explain the high observed Ly$\alpha$ equivalent widths. 
For the cosmological and star formation parameters
we adopted, the total stellar mass produced would seem to 
correspond to the formation of rather small galaxies, some of 
which are perhaps destined to merge. However, one of
our sources might become a serious candidate for a proto-giant
spheroidal galaxy if we assumed continuous star formation, a
low mass cutoff of 0.1 M$\odot$ in the IMF, and a 
flat accelerating universe with $\Omega_0$=0.2 and 
$\Omega_\Lambda$=0.8.

The implied star formation density in our sampled comoving volume 
is probably somewhat smaller than, but of the same order of 
magnitude as the star formation density at $z\sim3$
derived by other authors from Lyman-break galaxy surveys.
This result agrees with the expectation that 
the Ly$\alpha$ emitters are a low-metallicity (or low-dust)
tail in a distribution of star forming regions at high
redshifts. Finally, the Ly$\alpha$ emitters may contribute
as many H-ionizing photons as QSOs at 
$z\sim3$. They are therefore potentially significant for the 
ionization budget of the early universe. 
\end{abstract}


\section{Introduction}

Ly$\alpha$ emitting galaxies at high redshift are of much interest 
for studies of galaxy formation. While early surveys failed to
find such objects (e.g. Thompson et al. 1995), recent narrow-band 
imaging searches with spectroscopic follow-up identified a number of 
Ly$\alpha$ emitters at high $z$, first in fields near high-redshift
QSOs (Hu \& McMahon 1996) and later in blank fields (Cowie \& Hu 1998; 
Hu et al. 1998; Pascarelle et al. 1998). These Ly$\alpha$ 
emitters have typically very faint continua and high Ly$\alpha$
equivalent widths, i.e., may represent an early phase of galaxy
formation when substantial amounts of dust had not yet formed.

Here we report on the discovery of 9 Ly$\alpha$
emitters at redshift $z=3.1$ during our program of studying the
intracluster planetary nebula (PN) population in the Virgo cluster, 
which uses similar narrow-band imaging techniques to identify 
candidate PNs. The first indication of the existence of a diffuse 
intracluster stellar population in Virgo was the discovery by
Arnaboldi et al. (1996) that a few PNs in the galaxy NGC 4406 (M 86; 
redshift $-$230 km s$^{-1}$) have redshifts typical of the Virgo cluster 
(around +1300 km s$^{-1}$). Subsequently, direct evidence for red giant
stars belonging to this stellar population was reported by Ferguson et
al. (1998). Because PNs offer the chance to measure radial velocities
and perhaps even abundances for such a diffuse population, a search
for intracluster PNs in different positions across the Virgo cluster
started immediately and produced several dozens of PN candidates 
(M\'endez et al. 1997, Feldmeier et al. 1998) in a total surveyed 
area of $0.23$ deg$^2$.

The ``on-band/off-band'' narrow-band filter technique used to 
discover the PN candidates (see e.g. Jacoby et al. 1992) allows the
detection of a single emission line. This is not necessarily the
desired [O III] $\lambda$5007; it might be another emission
line at higher $z$, redshifted into the on-band filter. For example
[O II] $\lambda$3727 at $z=0.35$, or Ly$\alpha$ at $z=3.13$. In
previous work (M\'endez et al. 1997) we argued that most of our
detections had to be real PNs because (a) the surface density of
emission line galaxies derived from earlier studies was not high 
enough to explain all the detections; (b) the luminosity function
of the detected sources is in good agreement with the PN luminosity 
functions derived in several Virgo galaxies (Jacoby et al. 1990).

We have used the first ESO Very Large Telescope (VLT) unit (UT1) 
with the Focal Reducer and Spectrograph (FORS) in multi-object
spectroscopic mode, and the 4-m Anglo-Australian Telescope (AAT) 
with the 2-degree-field (2df) fiber spectrograph. Our purpose was
to confirm the nature of the Virgo intracluster PN candidates, 
to measure their radial velocities, and (in the VLT case) to detect 
the faint diagnostic lines required for abundance determinations. 

The present paper reports on the results of the VLT+FORS 
observations. The AAT+2df observations, which confirm the 
existence of Virgo intracluster PNs through the detection 
of both the $\lambda$4959 and $\lambda$5007 [O III] lines
(Freeman et al. 1999), will be presented and discussed by 
Freeman et al. (2000, in preparation).

\section{Observations and first results}

For the purpose of detecting faint nebular diagnostic lines with 
VLT+FORS (ESO programme 63.N-0530), we selected Field 1 of 
Feldmeier et al. (1998) as our first priority, because it had 
the brightest PN candidates, with a luminosity function cutoff
about half a magnitude brighter ($m_{5007} = 25.8$) than elsewhere 
in the Virgo cluster (see Feldmeier et al. 1998). The total area
of Field 1 is 256 arcmin$^2$. We also wanted to use FORS
(ESO programme 63.I-0007) to verify the PN nature of candidates in 
the smaller ``La Palma Field'' (50 arcmin$^2$, M\'endez et al. 
1997), with magnitudes 
$m_{5007}$ between 26.8 and 28.6. The lack of bright PNs in the 
La Palma Field was understandable in M\'endez et al. (1997) as a 
consequence of the sample size effect: if the total PN sample is 
small, the chance of finding a bright PN is too small. 

The observations were made with FORS at the VLT UT1 on the nights 
of 11/12, 15/16, 16/17 and 19/20 April 1999. Since the FORS field 
is slightly smaller than 7$\times$7 arc mins (with the standard 
collimator), we selected a portion of Field 1 where several
of the brightest candidates were located, and took on-band and 
off-band images of the selected portion of Field 1, and of the 
La Palma Field, on the night 11/12, using FORS in imaging mode.

The on-band and off-band interference filters we used have, 
respectively, central wavelengths 5039 and 5300 \AA, 
and FWHMs 52 and 250 \AA. Some 50 stellar images in 
the short on-band and off-band FORS exposures 
(10 min and 3 min, respectively) were used together with the 
corresponding stellar images in the Kitt Peak and La Palma
discovery images to define coordinate transformations that gave
the pixel values of the positions of the PN candidates in the
FORS images, given the pixel values of their positions in the
discovery images. In this way it was possible to define the FORS
slitlet positions with sufficient accuracy (in fact, the brightest 
PN candidates were visible in the short on-band FORS exposures,
allowing us to verify directly the accuracy of the pixel 
transformation, with typical errors below 0.5 pixels, which is
equivalent to 0.1 arc sec).

On the night of 15/16 April, R.P.K. and R.H.M. started FORS 
multi-object spectroscopy (MOS) in Field 1, using grism 300V 
without any order separation filter, for maximum spectral 
coverage. This grism gives a spectral resolution of about 
10 \AA, at 5000 \AA, for slitlets 1 arc sec wide. 
Of the 19 available slitlets, 10 were placed at the positions of 
PN candidates (numbers 1, 4, 5, 6, 16, 17, 21, 26, 27, 42, ordered 
by brightness as measured by Feldmeier). These candidates had 
magnitudes $m_{5007}$ between 25.8 and 26.4. The remaining 9 slitlets 
were placed on stars or galaxies in the field, in order to check the 
slitlet positioning (which was done by taking a short exposure 
without grism through the slitlets) and to help locate the dispersion 
lines as a function of position across the field, which is important
in the case of spectra consisting of isolated emission lines.

After taking 5 exposures of 40 min each in Field 1, all with the same 
slitlet configuration, and having made preliminary on-line reductions,
it was clear that the [O III] emission line candidates were not PNs. 
Object 1 showed one strong isolated emission line at 5021 \AA. 
It cannot be [O III] $\lambda$5007 because there is no hint of the 
companion line $\lambda$4959 at the corresponding wavelength. Object 5 
was identified as a starburst with $z=0.35$; it shows narrow emission
lines with a continuum, $\lambda$3727 is redshifted into the on-band 
filter, and H$\beta$ and the [O III] lines 4959 and 5007 are visible,
also redshifted with $z=0.35$. Object 17 was not detected. 
The remaining 7 objects were identified as continuum objects. 
Many of them are galaxies: they show redshifted emissions.

In summary, none of the candidates tested in Field 1 with VLT+FORS
are PNs. This result in fact solves a problem, because these Field 1 
candidates were surprisingly abundant and were somewhat brighter 
than typical PNs in the Virgo galaxies (see the discussion by 
Feldmeier et al. 1998). The high percentage of continuum objects 
(7 out of 9 detections) indicates that the off-band exposure by 
Feldmeier et al. was not deep enough. A re-examination of the 
Field 1 images by J. Feldmeier has subsequently confirmed
that due to sudden changes in the seeing and transparency the 
off-band does go to a brighter limiting magnitude than intended.
Briefly, the Field 1 exposures consisted of three 3600s on-band
exposures and five 600s off-band exposures taken at the Mayall 4-m
telescope.  The additional off-band exposures were intended 
to compensate for the transparency, which was decreasing 
at the time of the exposures.  Unfortunately, although these 
additional exposures do partially compensate for the change in 
transparency, the seeing increased by 0.3 arcsec, and consequently
the mean off-band exposure is not deep enough. 

On the other hand, Freeman et al. (1999) did confirm 
spectroscopically some fainter PN candidates in Field 1. 
Taken together, these results imply that the surface density 
of the intracluster stellar population originally estimated for
Field 1 has to be reduced (Freeman et al. 2000, in preparation). 
 
In view of the result in Field 1, we immediately pointed the VLT UT1 
to the La Palma field. In this field 12 PN candidates had been found,
11 reported in M\'endez et al. (1997) and one found afterwards.
The distribution of the 12 PN candidates in the sky made it impossible 
to assign slitlets to all of them in one slitlet configuration.
In our first configuration we defined 1 arc sec wide slitlets 
for 9 PN candidates and 2 objects suspected to be QSOs or starbursts
because they were visible (although much weaker) in the off-band
discovery image (M\'endez et al. 1997). On the April nights 15/16 
and 16/17 we completed 5 MOS exposures (40 min each) of the initial 
slitlet configuration, and on April 19/20 we took 3 additional MOS 
exposures of the La Palma Field (again 40 min each) with a different 
slitlet configuration, which allowed to add one PN candidate not 
observed before.

Thus spectra for a total of 10 PN candidates were acquired in the La 
Palma Field, and 7 were detected. They all show an isolated and 
narrow emission at wavelengths from 5007 to 5042 \AA. In all 
cases this is the only feature visible in the spectrum. 
The other 3 PN candidates, which are relatively faint, were not 
detected. Perhaps their slitlets were slightly misplaced, although 
we would think the errors in the pixel transformation were too small 
to have any effects on the detectability. Perhaps some of these
sources, if they are not PNs, have variable brightness.

Of the 2 QSO or starburst candidates, one was confirmed as a 
QSO at z=3.13, showing a typical broad-lined Ly$\alpha$ and C IV 
$\lambda$1550. The other one appears to be a starburst because it
shows one strong, isolated and narrow emission line, with faint 
continuum.

The success rate for emission line detection in the La Palma Field 
was satisfactorily high. It may be useful to give a few numbers 
for comparison with other searches, complementing information given
in M\'endez et al. (1997). The La Palma on-band image had a limiting
magnitude $m_{5007}$ = 28.7, equivalent to a flux of
10$^{-17}$ erg cm$^{-2}$ s$^{-1}$. 
We would collect the same amount of photons 
through the on-band filter from a star of visual magnitude 25.5. 
The off-band image had a limiting magnitude 0.2 mag fainter. 
The search for candidates was done by blinking the on-band versus 
the off-band image. To compare with the selection criteria used
e.g. by Steidel et al. (1999b) for their narrow-band imaging survey,
we define on-band and off-band magnitudes so that they coincide for 
an average star. All our emission-line candidates, being fainter or 
absent in the off-band image, have positive colors offband$-$onband. 
We can give only lower limits to the colors of candidates
undetected in the off-band image. All tested 
emission-line candidates with colors above 1.0 mag have been 
spectroscopically confirmed. The lower limits to the colors of 
the 3 objects that were not detected with FORS are below 1.0 mag.
Our 2 QSO or starburst candidates, detected in both images, 
have colors of 2.1 and 1.7 mag, respectively.

\section{Analysis of the La Palma Field spectra}

The CCD reductions were made using IRAF\footnote{IRAF is 
distributed by the National Optical Astronomical Observatories, 
operated by the Association of Universities for Research in 
Astronomy, Inc., under contract to the National Science 
Foundation of the U.S.A.} standard tasks. After bias 
subtraction, flat field correction and image combination to
eliminate cosmic ray events, the object spectra were extracted 
and the sky background subtracted. Then the He-Ar-Hg 
comparison spectra were extracted and the object spectra were 
wavelength calibrated. Spectrograms of the standard stars
G138-31 and G24-9 (Oke 1990) were used for the flux calibration.

We have designated the La Palma Field sources
with LPF plus a number ordered according to the brightness in the
discovery image. LPFnew is the latest PN candidate, found after the
discovery paper (M\'endez et al. 1997) was published. Object LPFs1 
is the QSO, and LPFs2 is the object identified 
as a starburst from the start, because of its stronger continuum.
F1-1 and F1-5 are objects 1 and 5 in Feldmeier's Field 1. 
F1-1 turns out to also have a very faint continuum, barely visible
in the final processed spectrogram.

Figs. 1 to 3 show, respectively, the spectra of the QSO, the 
starburst with visible continuum, and one of the La Palma Field
PN candidates showing no detectable continuum. All the LPF PN 
candidates look very similar, with only one emission line 
detected across the whole spectrum.

Having found no direct evidence of [O III] $\lambda$4959, which
would be expected if the detected emission line were 
$\lambda$5007, we can put an upper limit to the percentage of our 
emission-line objects that can be PNs. After rejecting objects 
that show a continuum, which clearly cannot be PNs, we proceed in 
the following way:

\noindent (1) shift the spectra, so that 
the wavelengths of the detected emission lines fall at 5007 \AA.

\noindent (2) normalize the intensities of the emission lines
to the same value, e.g. 300. 

\noindent (3) add all the spectra and measure the intensity of 
the resulting $\lambda$4959. If it is 50, for example, comparing 
to the expected value of 100 (since $\lambda$5007 was defined to 
be 300) we can argue that 50\% of the objects must be PNs. 

The result of this test is shown in Fig. 4, where we have
added the normalized spectra of the 7 PN candidates detected
in the La Palma Field. The complete absence of $\lambda$4959 
indicates that, at most, one of the 7 candidates can be a PN. 
This conclusion is based on the noise level, not on any marginal 
detection of $\lambda$4959. Besides, there is in Fig. 4 a hint of 
a weak continuum, not detectable in the individual spectra, which 
reinforces the rejection of these candidates as PNs. 

\section{Identification of the detected emission line as Ly$\alpha$}

Having rejected [O III]$\lambda$5007, because $\lambda$4959 is
not visible in Fig. 4, we consider the alternatives:

(1) [O II] $\lambda$3727 at $z=0.35$ was confirmed 
in one case (object 5 in Field 1) but can be rejected
in all other cases because we do not see H$\beta$, [O III]
$\lambda\lambda$4959, 5007 and H$\alpha$ at the corresponding 
redshifted wavelengths. This is illustrated in Figs. 5 and 6.

(2) Mg II $\lambda$2798 at $z=0.79$ can also be rejected 
for a similar reason: in this case we do not see [O II] 
$\lambda$3727 at the expected redshifted wavelength, as 
illustrated in Fig. 7. The same argument can be applied to other 
lines: assuming C III $\lambda$1909, we do not see $\lambda$2798; 
and so on.

We conclude that the isolated emission line must be Ly$\alpha$ 
at $z=3.1$. This identification is supported by the strength of 
the line: since we see at most a very faint continuum, the equivalent 
width is fairly large, typically 200 \AA \ (observed) and 50 \AA \ 
(rest frame). We have mentioned in M\'endez et al. (1997) that 
very few starburst galaxies show, for example, [O II] $\lambda$3727
stronger than 100 \AA \ in equivalent width.

\section{Implications for the surface density 
                         of intracluster PNs in Virgo}

A reliable estimate of the surface density of intracluster PNs in
Virgo will have to await a survey of a sufficiently large area on 
the sky, which is currently in progress.  Our results to date and
some preliminary conclusions are the following.

No PN candidates have been confirmed in the La Palma Field.
Only 5 of the 12 candidates have a chance of 
remaining as PNs: 2 were not tested and 3 were tested but not 
detected. Since we have identified most of these candidates as 
Ly$\alpha$ emitters, it is clear that the intracluster PN sample 
size in the La Palma Field and the inferred surface density must 
be substantially smaller than we estimated. 

On the other hand, the AAT+2df multi-object fiber spectroscopy has 
confirmed the existence of intracluster PNs in Fields 1 and 3 of 
Feldmeier et al. (1998). This will be reported in detail by Freeman
et al. (2000, in preparation). These PNs are brighter than 
$m_{5007} = 27$. Freeman et al. will show that the contamination by 
Ly$\alpha$ emitters at these brighter magnitudes is not as important
as in the La Palma Field. Thus it appears that the fraction of 
Ly$\alpha$ emitters in [O III] narrow-band selected samples is 
magnitude-dependent, increasing towards fainter values of $m_{5007}$.

The lack of bright PNs in the La Palma Field implies a
lower surface density than in other Virgo cluster positions, and
may indicate some degree of clustering in the distribution of the 
diffuse intracluster population. 

Our VLT+FORS observations
have shown that a spectroscopic confirmation of intracluster PN
candidates, involving the detection of both the $\lambda$4959 and
$\lambda$5007 [O III] emission lines, is necessary.

\section{Observed properties of the high-redshift 
                                        Ly$\alpha$ emitters}

In Table 1 we have collected some basic information about the 
narrow-lined sources: the measured Ly$\alpha$ fluxes, measured
Ly$\alpha$ wavelengths, redshifts, and Ly$\alpha$ equivalent widths
(W$_{\lambda}$ in almost all cases lower limits, because the 
continuum is not detected). The measured wavelengths provide yet 
another argument favoring the interpretation of all these
sources as unrelated to the Virgo cluster: they are randomly 
distributed across the on-band filter transmission curve, with no 
concentration at the Virgo cluster redshift (between 5020 and 
5030 \AA). See Fig. 8.

\begin{deluxetable}{lrccrrrr}
\tablecaption{Observed and derived properties 
                          of Ly$\alpha$ emitters. \label{tbl-1}}
\tablewidth{0pt}
\tablehead{
\colhead{object} & \colhead{flux\tablenotemark{a}} & 
\colhead{$\lambda$(\AA)} &
\colhead{$z$} & \colhead{W(\AA)\tablenotemark{b}} &
\colhead{M$_{\rm stars}$\tablenotemark{c}} &
\colhead{SFR\tablenotemark{d}} &
\colhead{M$_{\rm s}$\&SFR\tablenotemark{e}}}
\startdata
 LPFs2  & 17 & 5011 & 3.121 &        33 & 21-420 & 7-140 &   5-10\\
 LPF1   &  7 & 5011 & 3.121 &  $\geq$55 &   9-75 &  3-25 &   2-3\\
 LPF2   &  6 & 5026 & 3.133 &  $\geq$32 &  7-150 &  3-50 &   2-4\\
 LPF3   & 10 & 5042 & 3.146 & $\geq$175 &   8-11 &   3-4 &   -- \\
 LPF4   &  3 & 5007 & 3.118 &  $\geq$14 &  4-270 &  1-90 &   1-5\\
 LPF6   &  2 & 5035 & 3.140 &  $\geq$70 &   3-14 &   1-5 & 0.4-0.5\\
 LPF8   &  2 & 5034 & 3.140 &  $\geq$37 &   3-44 &  1-15 & 0.6-1\\
 LPFnew &  2 & 5010 & 3.120 &  $\geq$25 &   3-65 &  1-22 & 0.6-2\\
 F1-1   & 15 & 5021 & 3.129 &        45 & 19-200 &  6-67 &   4-6\\
\enddata
\tablenotetext{a}{measured Ly$\alpha$ fluxes in units of 
   10$^{-17}$ erg cm$^{-2}$ s$^{-1}$. Using the luminosity
   distance of 1.8 $\times 10^4$ Mpc, this column also gives the 
   Ly$\alpha$ luminosities, in units of 10$^8$ L$\odot$.}
\tablenotetext{b}{equivalent widths of Ly$\alpha$ transformed into
   the rest frame (the measured widths are $z+1$ times larger).}
\tablenotetext{c}{Possible range of total stellar mass formed, in units 
   of 10$^6$ M$\odot$, assuming starbursts of $3\times 10^6 {\rm yr}$
   duration; see section 7 and Fig. 14.}
\tablenotetext{d}{Possible range of star formation rates, in
   M$\odot$ yr$^{-1}$, assuming starbursts; see section 7.}
\tablenotetext{e}{Possible range of total stellar mass formed, in 
   units of 10$^9$ M$\odot$, assuming continuous star formation 
   over $10^9$ yr; see section 7 and Fig. 15. The same numbers also 
   represent the corresponding star formation rates in 
   M$\odot$ yr$^{-1}$.}
\end{deluxetable}

Notice that object LPF3 is brighter than measured in the discovery 
image. The reason is that in this case Ly$\alpha$ falls near the 
edge of the on-band filter transmission curve. LPF3 is remarkable
also for the very large lower limit to its Ly$\alpha$ equivalent width.

The observed fluxes are between $2 \times 10^{-17}$ and 
$2 \times 10^{-16}$ erg cm$^{-2}$ s$^{-1}$. 
We are looking into a small redshift range, defined by the 
transmission of the on-band filter used at La Palma, of about 
$\Delta{\rm z} = 0.04$. The La Palma Field covers 50 arcmin$^2$.
The total Ly$\alpha$ flux measured within our ``discovery box''
is $5 \times 10^{-16}$ erg cm$^{-2}$ s$^{-1}$.
Adopting $z=3.13$, H$_0$=70 km s$^{-1}$ Mpc$^{-1}$ and q$_0$=0.5
we get a luminosity distance of 1.8 $\times 10^4$ Mpc, which implies 
Ly$\alpha$ luminosities for our sources between $2 \times 10^8$ and 
$2 \times 10^9$ L$\odot$. The total Ly$\alpha$ luminosity within the 
sampled volume is $5 \times 10^9$ L$\odot$.
The sampled comoving volume is 1650 Mpc$^3$, which gives from 8 
sources a comoving space density $5 \times 10^{-3}$ Mpc$^{-3}$.

These numbers depend on our assumptions 
about the cosmological parameters: for a flat universe with 
$\Omega_0$=0.2, $\Omega_\Lambda$=0.8, and the same $z$ and H$_0$ 
as above, the luminosity distance becomes 3 $\times$ 10$^4$ Mpc, 
the sampled comoving volume 10$^4$ Mpc$^3$, and the Ly$\alpha$ 
luminosities become larger by a factor 2.8.

\section{Starburst models and derived quantities}

What produces the narrow Ly$\alpha$ emission? Given rest frame 
equivalent widths below 200 \AA, the most probable source of 
ionization is massive star formation (Charlot and Fall 1993). 
Active galactic nuclei (AGNs) might be an alternative, although 
there is no hint of C IV 1550 in our spectra, as shown in Fig. 9.
We have taken massive star formation as our working hypothesis.

We have made a simple population synthesis model to explore
some basic properties of the stellar population responsible for 
the ionization of the H II regions. 
Our main interest in doing this analysis is to estimate star
formation rates and densities; we would like to verify if these
Ly$\alpha$ sources correspond to the formation of massive giant
galaxies or rather to the formation of smaller structures. 

We adopt a standard initial mass function (IMF), 
f(M) $\propto $ M$^{-2.35}$, stellar evolutionary models for 
low metallicity (Z=0.001; this choice of metallicity
will be justified in next paragraph) from Schaller et al. (1992), 
and ionizing fluxes from NLTE model atmospheres with winds 
(Pauldrach et al. 1998), again for a low metallicity, in this case 
that of the SMC (5 times below solar).

The relation between the number of stellar Lyman continuum photons
$N_{LyC}$ and the Ly$\alpha$ luminosity can be obtained from a 
simple recombination model:

\begin{equation}
{\rm L}({\rm Ly}\alpha) = {\rm h} \nu({\rm Ly}\alpha) 
                                            \ 0.68 \ X_B \ N_{LyC}
\end{equation}

\noindent where 0.68 is the fraction of recombinations that yield 
Ly $\alpha$ (Case B, see e.g. Storey and Hummer 1995), 
and $X_B$ is the product of the fraction of $N_{LyC}$ 
really absorbed times the fraction of Ly$\alpha$ photons that 
really escape. We necessarily have $0 \leq X_B \leq 1$.
Note that $X_B \sim 1$ requires both an optically thick nebula
and very low dust content, because the large number of Ly$\alpha$
scatterings in an optically thick nebula will lead to their 
absorption by dust grains, if such grains are present in any 
significant number. This explains our choice of a low metallicity.
How low is low? Since we do not know much about dust properties,
we have used empirical information provided by Charlot and Fall
(1993) in their Fig. 8, which shows Ly$\alpha$ equivalent widths
as a function of oxygen abundance in nearby star-forming galaxies.
In that figure we find that Ly$\alpha$ equivalent widths larger
than 20 and 50 \AA \ are associated respectively with oxygen 
abundances below 25\% and 10\% solar. In future work, to obtain
more quantitative constraints, we intend to carry out Ly$\alpha$
radiative transfer calculations in the presence of dust and
velocity fields as an extension of the work by Hummer and Kunasz 
(1980), Hummer and Storey (1992) and Neufeld (1990, 1991).

Figs. 10 and 11 show Ly$\alpha$ equivalent widths 
for single stars as a function of $X_B$ and of the stellar 
T$_{\rm eff}$. We need
both a large $X_B$ and many massive, hot main sequence stars 
in order to produce the observed Ly$\alpha$ equivalent widths.

We have explored two different histories of star formation in a
low-metallicity stellar population: (a) a starburst, and 
(b) continuous star formation. These two alternatives are defined
as star formation extending over a time (a) similar to 
($3 \times 10^6$ years) and (b) much longer than (10$^9$ years) 
the duration of a main sequence OB star. 

Figs. 12 and 13 show the resulting contour plots of Ly$\alpha$
equivalent widths as a function of $X_B$ and of the maximum 
main sequence mass in the population. In these figures, the
quantities on the x axis allow different interpretations. 
In the case of a starburst, smaller values of $M_{\rm max}$
can be interpreted to  
represent an increasing age of the starburst, or in other words 
the time elapsed since the starburst happened; as the starburst 
grows older, the most massive stars are removed from the main 
sequence. In the case of continuous star formation, the value 
of $M_{\rm max}$ indicates at which mass the integration of the IMF 
was stopped. In other words, the left part of the plot shows a case
in which very massive main sequence stars were not formed.

We can obtain Ly$\alpha$ equivalent widths $ > 50$ \AA \
only for $X_B > 0.5$ (continuous star formation) or $X_B > 0.3$ 
(starburst). This points to almost completely optically thick,
extremely dust-poor nebulae. 
It is easy to understand why a large $X_B$ is necessary. If it 
is small, many stellar ionizing photons are lost. In order to 
explain the observed Ly$\alpha$
fluxes we must add more massive stars, but these stars make
a strong contribution to the continuum, and therefore the 
Ly$\alpha$ equivalent width must decrease.

It may seem surprising that we can get large Ly$\alpha$ equivalent 
widths at so low values of $M_{\rm max}$. The reason is that the
Schaller et al. main sequence at low metallicity is shifted to 
rather high T$_{\rm eff}$ because of the low stellar opacity. This
means that lower mass objects on the main sequence give much more 
ionizing flux than at solar metallicity.

The source LPF3 can be explained only as a very young starburst 
with $X_B \sim 1$, because of its very large Ly$\alpha$ equivalent 
width.

We set the lower mass limit of the Salpeter IMF at 0.5 M$\odot$. 
For a maximum stellar mass of 120 M$\odot$, this gives an average 
mass of 1.65 M$\odot$, which is then the conversion factor 
between number of stars and total stellar mass. The average 
mass decreases only slightly if we decrease $M_{\rm max}$,
and only if $M_{\rm max}$ is interpreted as in Fig. 13, because
in that case very massive stars are not formed. The average mass
does not decrease if $M_{\rm max}$ is low due to the age of the
starburst, because the very massive stars are assumed to have 
formed and evolved away from the main sequence.

Figs. 14 and 15 show contour plots of the logarithms of Ly$\alpha$ 
luminosities in a plane where the x axis is the same $M_{\rm max}$ 
used in Figs. 12 and 13, and the y axis is the product of $X_B$ 
times the total number of stars produced over the total duration
of the star formation ($3\times 10^6$ and $10^9$ yr, respectively). 
The number of stars required to
produce a given Ly$\alpha$ luminosity depends on the value of 
$M_{\rm max}$. Since we cannot determine $M_{\rm max}$ empirically
or derive it from first principles, we have considered the full
range of $M_{\rm max}$, from 120 M$\odot$ to as low as permitted
by the limits on the Ly$\alpha$ equivalent widths; sometimes as
low as 20 M$\odot$. This produces rather large uncertainties in
the derived masses, particularly in the case of starbursts.
A comparison of Figs. 14 and 15 also shows that, as expected, 
in the case of continuous star formation many more stars 
are needed to obtain a given Ly$\alpha$ luminosity. 

The results of these mass estimates are listed in Table 1.
The total masses of stars and star formation rates 
turn out to be very sensitive to the star formation history.
For continuous star formation and young starbursts, i.e. high
values of $M_{\rm max}$, our strongest sources have
SFRs of the order of 10 M$\odot$ yr$^{-1}$. However these
numbers increase dramatically if we consider older starbursts,
and we cannot rule out SFRs higher than 200 M$\odot$ yr$^{-1}$
in a few cases. 

Adding up the masses and SFRs in Table 1, and using the average 
stellar mass of 1.65 M$\odot$, we have built Table 2, which 
provides the total number and total mass of stars formed,
and the resulting 
star formation rate, needed to explain our total Ly$\alpha$ 
luminosity of $5 \times 10^9$ L$\odot$ in the La Palma Field,
for the cases of starbursts and continuous star formation.
Here we have not added the numbers for F1-1, which belongs to 
another field.

\begin{deluxetable}{lrrr}
\tablecaption{Star formation needed to explain the total Ly$\alpha$ 
                      luminosity in the La Palma Field. \label{tbl-2}}
\tablewidth{0pt}
\tablehead{
\colhead{star formation} & \colhead{N$_{\rm stars}$\tablenotemark{a}} & 
\colhead{total M$_{\rm stars}$\tablenotemark{b}} &
\colhead{SFR\tablenotemark{c}}}
\startdata
   starburst  & 4-65 $\times 10^7$  & 6-105 $\times 10^7$ & 19-350 \\
   continuous & 7-15 $\times 10^9$  & 12-26 $\times 10^9$ & 15-30 \\
\enddata
\tablenotetext{a}{Total number of stars formed}
\tablenotetext{b}{Total mass of stars formed, in M$\odot$}
\tablenotetext{c}{Star formation rate in M$\odot$ yr$^{-1}$}
\end{deluxetable}

How uncertain are these numbers? First of all, the real SFR might
be higher than we inferred for our sources because of extinction
by dust. However, we consider this to be unlikely because it would
be necessary to argue that there is dust in the foreground (to produce
extinction) but not inside (it would destroy the Ly$\alpha$ emission).
We do not argue in terms of low metallicity because there is at least
one case of a very metal-deficient blue compact dwarf galaxy, 
SBS 0335-052, where dust patches are clearly visible (Thuan and 
Izotov 1999). Note however that this source has Ly$\alpha$ in
absorption, implying that extinction is accompanied by destruction
of Ly$\alpha$ emission, as expected. Based on this reasoning we do 
not include a correction for extinction in our Tables.

The reader may argue that geometry probably plays the dominant role 
in dictating how many Ly$\alpha$ photons escape, making our attempt
to estimate star formation rates a futile exercise. However, our
sources are a rather special case. Geometry is useful when we need
to explain why a star-forming region shows Ly$\alpha$ in absorption;
see e.g. Kunth et al. (1998). But our sources do show strong Ly$\alpha$
emission, and furthermore, show a very faint continuum. The 
absence of continuum (i.e. the large equivalent width of Ly$\alpha$)
is very important because it precludes the existence of large amounts 
of hot stars. If there is any significant destruction of Ly$\alpha$
photons, then, in order to explain the observed Ly$\alpha$ fluxes we
must add more massive stars. But these additional stars contribute to 
the continuum, and the Ly$\alpha$ equivalent width decreases too much.
Thus the absence of continuum acts as a "safety valve" restricting
higher values of the star formation rate. As explained above,
extinction by dust is not very likely.

On the other hand, the derived masses and SFRs are strongly dependent 
on the lower mass limit of the IMF. If we selected a lower mass 
limit of 0.1 instead of 0.5 M$\odot$ there would be
no change in the number of stars needed to produce the ionizing 
photons, but we would be adding a lot of low-mass stars; 
the total mass estimate for a given total Ly$\alpha$ luminosity 
would increase by a factor 1.9. The SFRs would have to be 
corrected by the same factor. The ranges of masses and SFRs in 
Tables 1 and 2 do not include this source of uncertainty, but we 
will take it into account in the discussion.

The adopted cosmological parameters also have some influence.
We have adopted H$_0$=70 km s$^{-1}$ Mpc$^{-1}$ and q$_0$=0.5, 
which implies a sampled comoving volume of 1650 Mpc$^3$.
If we take, for example, a total SFR of 50 M$\odot$ yr$^{-1}$, 
we get a star formation density in the comoving volume of 0.03 
M$\odot$ yr$^{-1}$ Mpc$^{-3}$.
If we adopted a universe with $\Omega_0$=0.2, $\Omega_\Lambda$=0.8, 
but with the same $z$ and H$_0$, then our star masses and SFRs would 
be larger by the same factor 2.8 as for the luminosities. However the 
star formation density mentioned above would drop from 0.03 to 0.014
M$\odot$ yr$^{-1}$ Mpc$^{-3}$ because of the enlarged sampled 
comoving volume of 10$^4$ Mpc$^3$.

\section{Discussion}

Recent blank field searches for Ly$\alpha$ emitters 
have been successful at redshifts from 2.4 to 5: Hu 1998, 
Cowie and Hu 1998, Pascarelle et al. 1998, Hu et al. 1998.
The sources that have been verified spectroscopically 
(Hu 1998, Hu et al. 1998) are like ours: strong, 
narrow, isolated emission line identified as 
Ly$\alpha$, with a similar range of equivalent widths. 

One may ask why the earlier searches failed to detect such 
emission-line sources. For example Thompson et al. (1995) 
reported no detection in a total area of 180 arcmin$^2$
down to a 1-$\sigma$ limiting flux 10 times fainter than 
the flux of our brightest source, LPFs2. This might be 
attributed to a very low surface or space density in their 
directions, but since they looked in many directions this 
interpretation looks improbable. The surface density in
the La Palma Field does not seem to be abnormally high.
The 8 sources we detected in the 50 arcmin$^2$ 
of the La Palma Field and within our redshift range of 0.04 
are equivalent to 14400 emitters deg$^{-2}$ per unit $z$ 
with Ly$\alpha$ fluxes above 
$1.5 \times 10^{-17}$ erg cm$^{-2}$ s$^{-1}$. This is a lower
limit to the true density of such sources; if some of the three
photometric candidates not detected and of the two candidates not
tested are Ly$\alpha$ emitters with similar fluxes, the true density
could be higher by up to 60\%. The density inferred from the present
spectroscopic sample is similar to the 15000 deg$^{-2}$ per unit $z$ 
reported by Hu et al. (1998) from Keck searches in 
different regions of the sky, with the same limiting flux.
Our sources are brighter than those detected with the Keck 
telescope (Hu 1998, Cowie and Hu 1998, Hu et al. 1998),
but our redshift is smaller; see Fig. 16. In summary, the  
surface density of Ly$\alpha$ emitters in the La Palma Field 
is of the same order of magnitude as for the sources detected
by Hu et al. Pascarelle et al. (1998) report 
an order-of-magnitude range of space densities at $z=2.4$, in 
some cases even higher than ours. More surveys will probably
clarify whether or not there is any structure or clustering  
in the distribution of Ly$\alpha$ emitters.

Concerning masses and star formation rates, if we adopt continuous 
star formation or young starbursts, where star formation has not yet 
(or has just) stopped, then from Table 1 we read that our strongest 
sources show a SFR not higher than about 10 M$\odot$ yr$^{-1}$, 
similar to the maximum
SFRs reported by Hu et al. (1998), and short of what would be 
expected from a proto-giant spheroidal galaxy forming more than 
10$^{11}$ M$\odot$ in 10$^{9}$ years. We cannot completely rule out 
the possible existence of an inconspicuous stellar population,
already formed, which would be detectable only in the infrared, 
but we would find it difficult to explain the very low dust content
in that case. The possible existence of undetected neutral and 
molecular H gas would provide additional means to scatter and
eventually destroy the Ly$\alpha$ photons (e.g. Neufeld 1990), and 
therefore we consider 
it unlikely. Thus from the available evidence we would seem to be 
witnessing the formation of small subgalaxies, some of which are
perhaps destined to merge. This conclusion is also supported 
by the total mass produced, even in the case of continuous star 
formation, and would be unchanged if we adopted the lowest IMF mass 
limit of 0.1 M$\odot$. If we also adopted the flat accelerating 
universe with $\Omega_0$=0.2 and $\Omega_\Lambda$=0.8, which 
implies larger distances, luminosities and SFRs, the starbursts 
would still point to small entities, but 
the continuous star formation in the case of LPFs2 would be able
to produce several times 10$^{10}$ M$\odot$. 

On the other hand, continuous star formation implies the previous
production of supernovae and metals, which makes it more difficult
to explain a very low dust content. For that reason we consider 
young starbursts more likely than continuous star formation or 
old starbursts. But in dealing with this subject we prefer to
be cautious and leave all conceivable options open.

What would happen if we allowed for older starbursts, in which
star formation has stopped 10$^7$ years ago? Then the mass of stars
formed would increase substantially, but notice that in such cases
the mass produced after star formation has stopped is not larger 
than 10$^9$ M$\odot$. Therefore even in this extreme case 
starbursts are not able to make a proto-giant spheroidal galaxy
out of any of our sources. This conclusion would not be weakened 
by assuming the accelerating universe mentioned above.

Now we try to estimate lower and upper limits for the star formation 
density in the sampled comoving volume that corresponds to the La 
Palma Field, assuming that we can rule out production of 
Ly$\alpha$ photons through AGN activity. A lower limit for the
SFR of 15 M$\odot$ yr$^{-1}$ comes from the continuous star formation 
case in Table 2. The maximum SFR in Table 2 (old starbursts)
is rather improbable, because it is obtained assuming that all the
starbursts ended at the same time, some 10$^7$ years before the
Ly$\alpha$ photons we detected were emitted. A more reasonable upper 
limit is obtained assuming that one of the starbursts is at the 
right age to produce the maximum SFR, while all 
the others make smaller contributions. 
Let us assume that the source produced by the only 
allowed old starburst is LPFs2, which would then
contribute 140 M$\odot$ yr$^{-1}$. Adding the almost negligible
contributions from the other sources, we obtain in this way a range 
of plausible total SFRs in the comoving volume between 15 and 170 
M$\odot$ yr$^{-1}$. Allowing now for the uncertainty in the lower
mass limit of the IMF, we get a range between 15 and 320 
M$\odot$ yr$^{-1}$, which implies a star formation density between 
0.009 and 0.19 M$\odot$ yr$^{-1}$ Mpc$^{-3}$. For comparison,
Hu et al. (1998) obtained, from their own samples, 0.01 in the 
same units.

In order to compare our star formation density with other 
available information, we adopt in this paragraph 
H$_0$=50 km s$^{-1}$ Mpc$^{-1}$ and q$_0$=0.5, which is what
other authors have done, e.g. Steidel et al. (1999a).
This gives luminosities twice as large as in our earlier choice,
and a sampled comoving volume of 4500 Mpc$^3$. Our resultant
star formation density drops slightly to between 0.007 and 0.14 
M$\odot$ yr$^{-1}$ Mpc$^{-3}$.
This is plotted in Fig. 17 together with other data collected 
from several galaxy surveys by Steidel et al. (1999a).
Our star formation density in Ly$\alpha$ emitters is comparable to
that in other star forming sources at that redshift. Our sources are
probably the low-metallicity (or low-dust) tail in a distribution of
star forming regions at high redshifts.  This is underlined by the
fact that in our sampled volume we also have a QSO (LPFs1) with 
apparently higher metal abundances, judging from the strength of 
the C, N, O lines in its spectrum.

As already remarked by Hu et al. (1998), since we expect
lower metallicity at higher redshifts, we should expect the strong 
Ly$\alpha$ emitters to become more frequent at higher redshifts 
relative to Lyman-break galaxies, which normally have weak or 
absent Ly$\alpha$ emission and are therefore presumably more 
metal-rich. Note, however, that low metallicity does not necessarily
imply emission in Ly$\alpha$: the blue compact dwarf galaxy 
SBS 0335-052 (Thuan and Izotov 1999), with $Z=Z_{\odot}/41$ and 
Ly$\alpha$ in absorption, provides a beautiful cautionary note. 
Kunth et al. (1998) have argued that the geometry and velocity 
structure of the interstellar medium play an important role in 
determining the strength of the Ly$\alpha$ emission, and this may 
complicate the comparison between Ly$\alpha$ emitters and 
Lyman-break galaxies.

Finally, we consider our Ly$\alpha$ emitters as possible sources
of ionization of the intergalactic medium at high redshift
(see e.g. Madau et al. 1998). Keeping the same cosmological
parameters as in last paragraph, we get a total L(Ly$\alpha$),
in our sampled comoving volume, of $10^{10}$ L$\odot$. Converting 
this into $N_{LyC}$ using Eq. (1), on the assumption that 
$X_B = 0.5$, we get $7 \times 10^{54}$ photons s$^{-1}$. Since the
sampled comoving volume is 4500 Mpc$^3$, this implies a ionizing 
photon density of 1.5 $\times 10^{51}$ photons s$^{-1}$ Mpc$^{-3}$.
This is the total number of ionizing photons produced by the stars.
It is very difficult to estimate what fraction is available for
ionization of the intergalactic medium, because it depends on the
relative contributions of the two factors that enter into $X_B$.
In at least one case (LPF3) we have $X_B \sim 1$, which
means that this source cannot contribute many ionizing photons.
Assuming optimistically that there remain $10^{51}$ ionizing
photons s$^{-1}$ Mpc$^{-3}$ available, our Ly$\alpha$ emitters 
seem to contribute about 0.5 times as much as the 
star-forming galaxies at $z=3$ found in Lyman-break galaxy
surveys (see e.g. Fig. 2 of Madau et al. 1998). 
We conclude that the contribution by Ly$\alpha$
emitters may be comparable in order of 
magnitude to what QSOs provide at $z=3$. 

\section{Summary of conclusions and perspectives}

We have discovered a population of high-redshift Ly$\alpha$
emitters in our sample of Virgo intracluster PN candidates obtained
with an ``on-band/off-band'' filter technique. Our VLT+FORS spectra 
show that the Ly$\alpha$ emitters at $z=3.13$ 
look very similar to those discovered in other fields 
at other redshifts by Hu et al. (1998). Only a narrow 
and strong Ly$\alpha$ emission is visible in their spectra; 
no other spectral line, and no (or at most a very weak) 
continuum. On the assumption that Ly$\alpha$ emission is 
produced by massive star formation, we have estimated 
the total mass of stars formed and star formation
rates, and we have estimated
the star formation density in our sampled comoving volume.
The Ly$\alpha$ emitting nebulae must be nearly optically thick 
and extremely dust-poor, probably indicating a very low 
metallicity. The total mass formed and the SFRs appear to 
suggest that we are witnessing the formation of rather small
galaxies. This conclusion depends on several assumptions
about the IMF, star formation history and cosmological 
parameters. There is one source (LPFs2) that might qualify 
as a proto-giant spheroidal if we assumed continuous star 
formation, a lower mass cutoff of 0.1 M$\odot$ in the IMF, 
and a flat accelerating universe with $\Omega_0$=0.2 and 
$\Omega_\Lambda$=0.8. However, to assume continuous star 
formation 
implies the previous production of supernovae and metals,
making it more difficult to explain the very low dust content
required by the observed spectra of our sources. We are more
probably dealing with young starbursts.

Taking all sources into account, the implied star formation 
density in our sampled comoving volume is probably 
somewhat smaller than, but of the same order of 
magnitude as the star formation density at $z\sim3$
derived by other authors from Lyman-break galaxy surveys.
This result agrees with the expectation that 
the Ly$\alpha$ emitters are a low-metallicity (or low-dust)
tail in a distribution of star forming regions at high
redshifts. Finally, the Ly$\alpha$ emitters may contribute
as many H-ionizing photons as QSOs at 
$z\sim3$. They are therefore potentially significant for the 
ionization budget of the early universe. 

More extensive surveys at different redshifts will be needed to build 
a luminosity function for the Ly$\alpha$ emitters and to decide if 
they show any evidence of clustering and evolution as a function of 
redshift. HST images might be able to resolve these sources; 
their morphology might offer clues about their star formation 
processes. High-resolution spectroscopy of Ly$\alpha$ would
provide important kinematic information about the interstellar 
medium in these galaxies. We need infrared spectra to try to 
detect forbidden lines and either determine or put some upper limits 
to the metallicity. It might also be possible to clarify to what 
extent the production of Ly$\alpha$ photons can be attributed to AGN 
activity. The firm detection of an infrared continuum would help to 
constrain the characteristics and total mass of the stellar 
populations through comparison with population synthesis models.

The implications of our results for the surface density of the 
diffuse intracluster stellar population in the Virgo cluster are not 
clear yet. Future searches for intracluster PNs will have to include the
spectroscopic confirmation of the candidates, by detection of the
two bright [O III] emissions at 4959 and 5007 \AA. From the
spectroscopic work of our group to date, including the confirmation of
23 intracluster PNs by the detection of the two [OIII] lines
by Freeman et al.\ (1999), the fraction of
high-redshift Ly$\alpha$ sources in the on-band/off-band samples
appears to be higher at faint magnitudes. In the specific case of the
``La Palma Field'' we have not found any intracluster PNs, perhaps
suggesting some degree of clumpiness in the distribution of the
intracluster stellar population. In other Virgo fields with
brighter PN candidates, the fraction of high-redshift Ly$\alpha$
emitters among the detected sources appears to be about 25\% (see
Freeman et al. 2000, in preparation, for a more extensive discussion).

\acknowledgments

RPK would like to thank the director and staff of Steward Observatory,
Tucson, for their hospitality and support during an inspiring sabbatical
which among other results has led to the ESO proposal 63.N-0530.
Thanks also to Stefan Wagner for helpful comments.
RPK and RHM express their gratitude to the ESO staff at Cerro Paranal,
who contributed with their efforts to make this VLT run an enjoyable 
experience.

\newpage

\notetoeditor{Figs. 1, 5, 9 are intended for the full width of the
page (two columns). All the other figures fit in one column.}

\figcaption[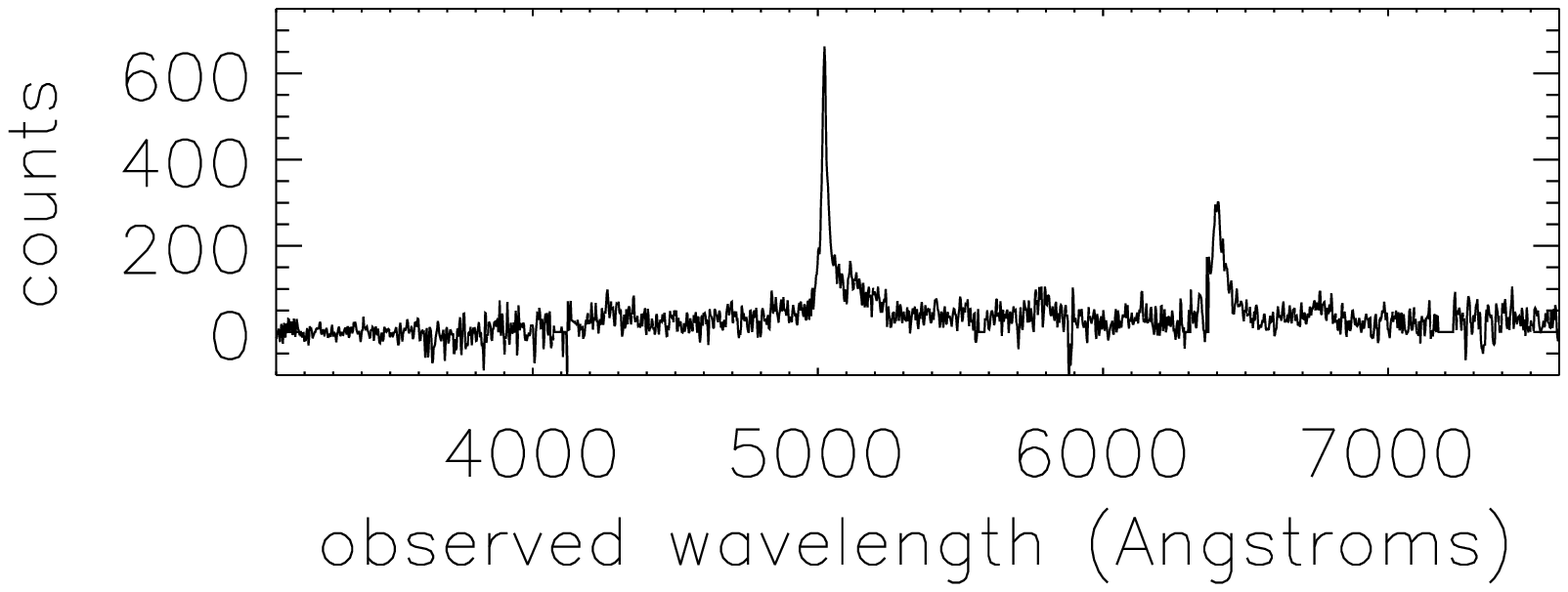]{The spectrum of QSO candidate LPFs1. The 
strong, broad emissions (Ly$\alpha$ + N V $\lambda$1240 and 
C IV $\lambda$1550) confirm that this is a QSO at a redshift 
$z=3.13$ (the observed wavelength of Ly$\alpha$ is 5024 \AA).
A faint continuum is visible only longward of 3760 \AA, which
is consistent with the position of the Lyman break. Other features 
visible are O VI $\lambda$1035 at 4275 \AA, O IV + Si IV 
$\lambda$1402 at 5790 \AA, and He II $\lambda$1640 at 6775 \AA.
\label{fig1}}

\figcaption[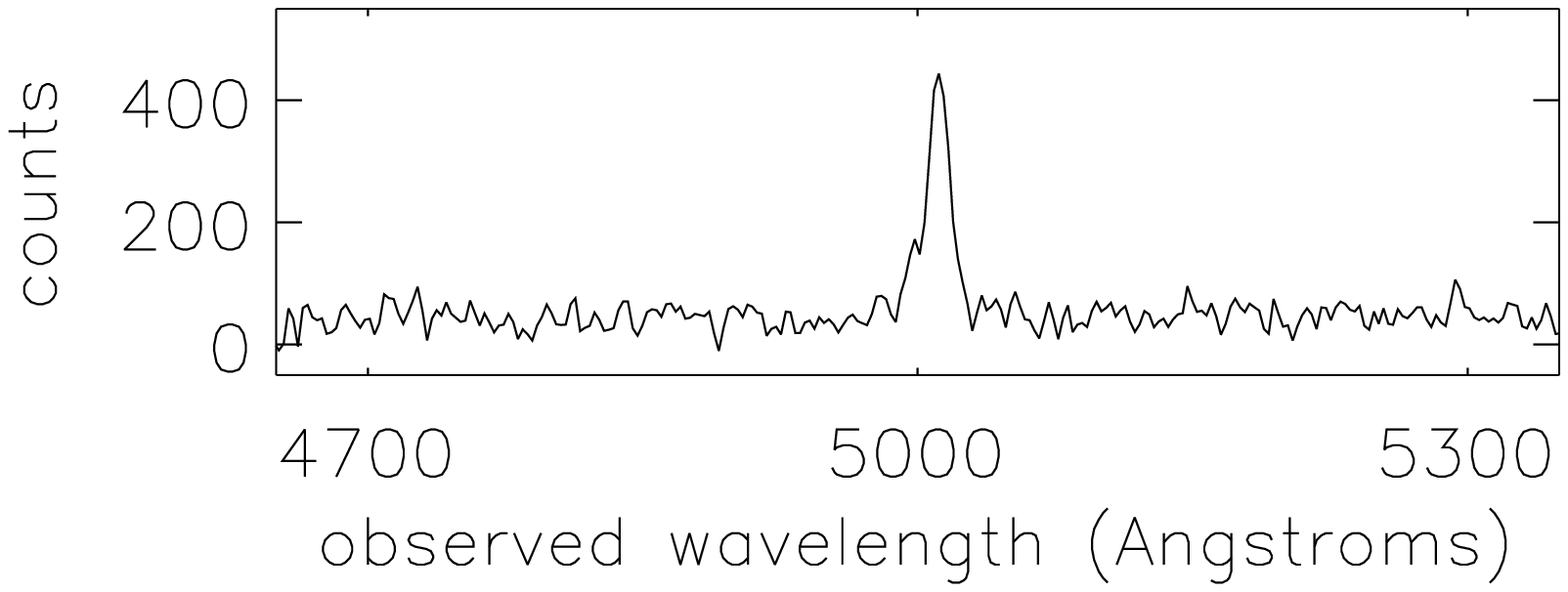]{The isolated emission in the spectrum of
starburst candidate LPFs2. A continuum is clearly seen, as expected 
from the off-band detection in the discovery paper. \label{fig2}}

\figcaption[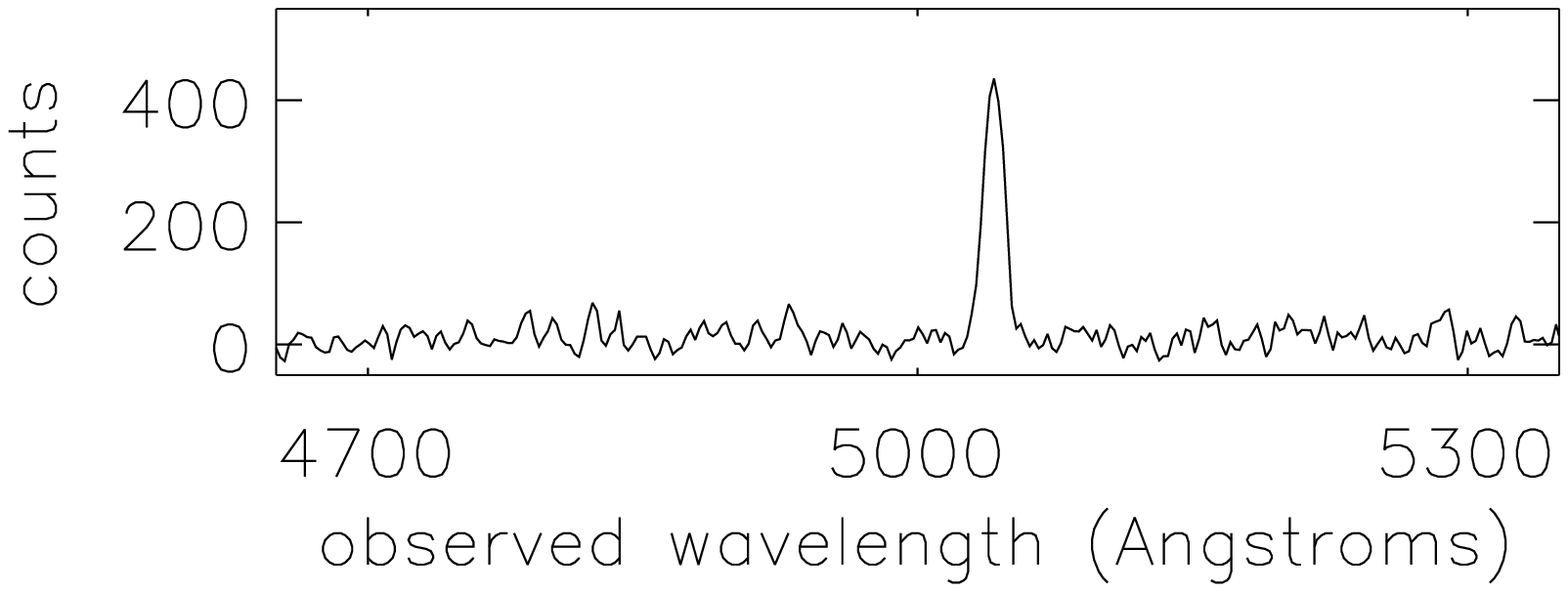]{The isolated emission in the spectrum of 
PN candidate LPF3. Here no continuum is detected, consistent with 
the off-band non-detection in the discovery paper. \label{fig3}}

\figcaption[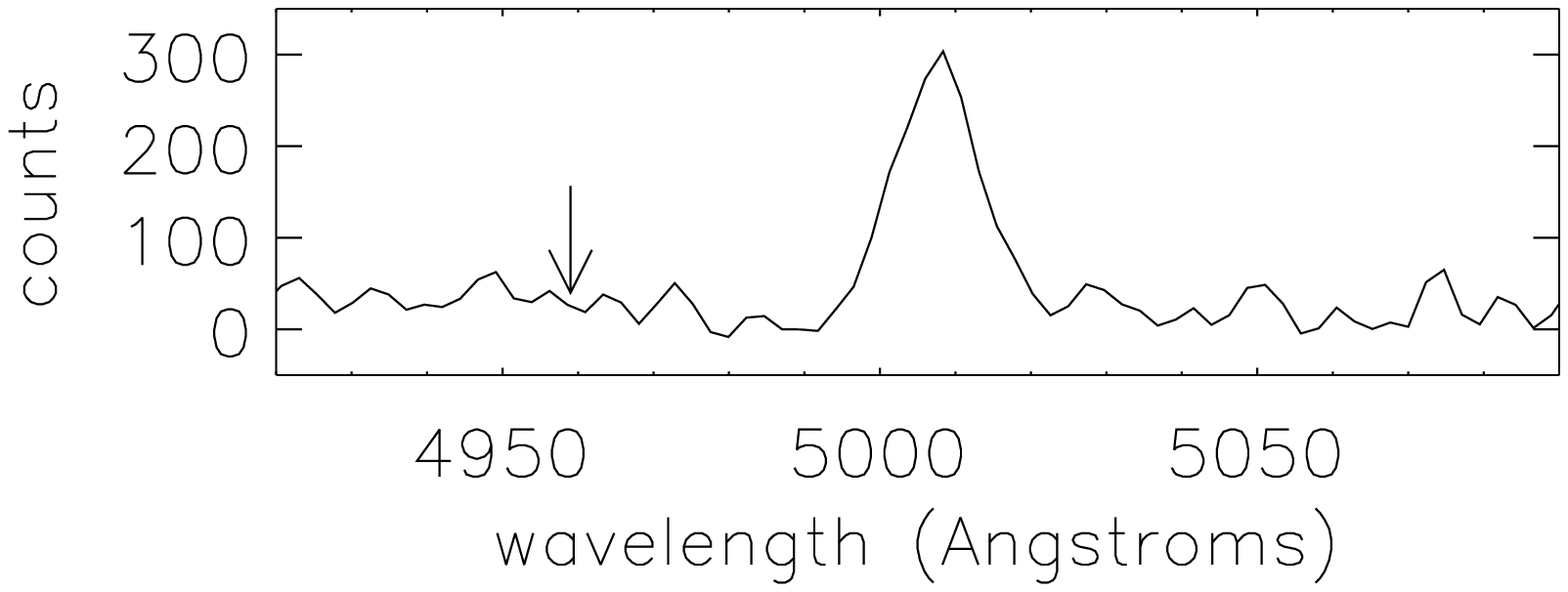]{Addition of the normalized spectra of 7 PN
candidates (no individual continua detected). The normalization
procedure is described in section 3. The arrow indicates where
[O III] $\lambda$4959 should be, if the visible emission were
[O III] $\lambda$5007. The absence of 4959 leads to conclude that
probably none, and at most one, of the 7 candidates can be a PN. 
Note also that 
in this added spectrum a very faint continuum may be visible. 
\label{fig4}}

\figcaption[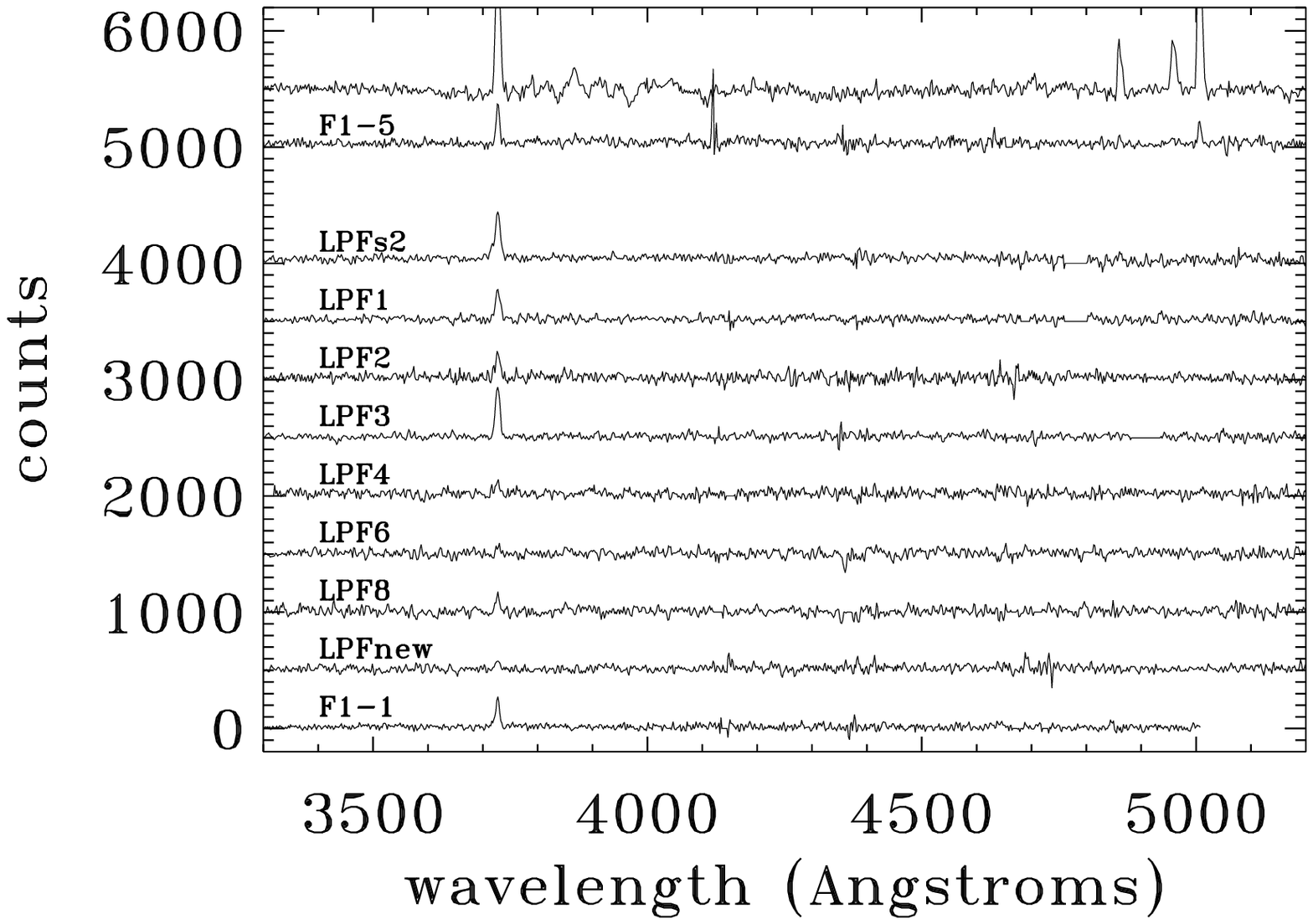]{The upper spectrum corresponds to an anonymous
 galaxy in the La Palma Field which happened to show strong 
emission at [O II] $\lambda$3727, 
H$\beta$ and [O III] $\lambda\lambda$4959, 5007. We
use this object as reference; its continuum has been arbitrarily
rectified for easier comparison. Below this star-forming galaxy
we show the spectrum of F1-5 (object 5 in Field 1), 
redshift-corrected so that the strong emission line detected with 
the on-band filter falls at 3727 \AA. F1-5 shows the same set of 
starburst emission lines as the reference galaxy, and therefore its 
redshift is confirmed to be 0.35. Below we have plotted the spectra 
of our sources, all redshift-corrected in the same way as
for F1-5. The levels of zero intensity are separated by 500 
counts. None of these sources
shows other emission lines at the relevant wavelengths. 
\label{fig5}}

\figcaption[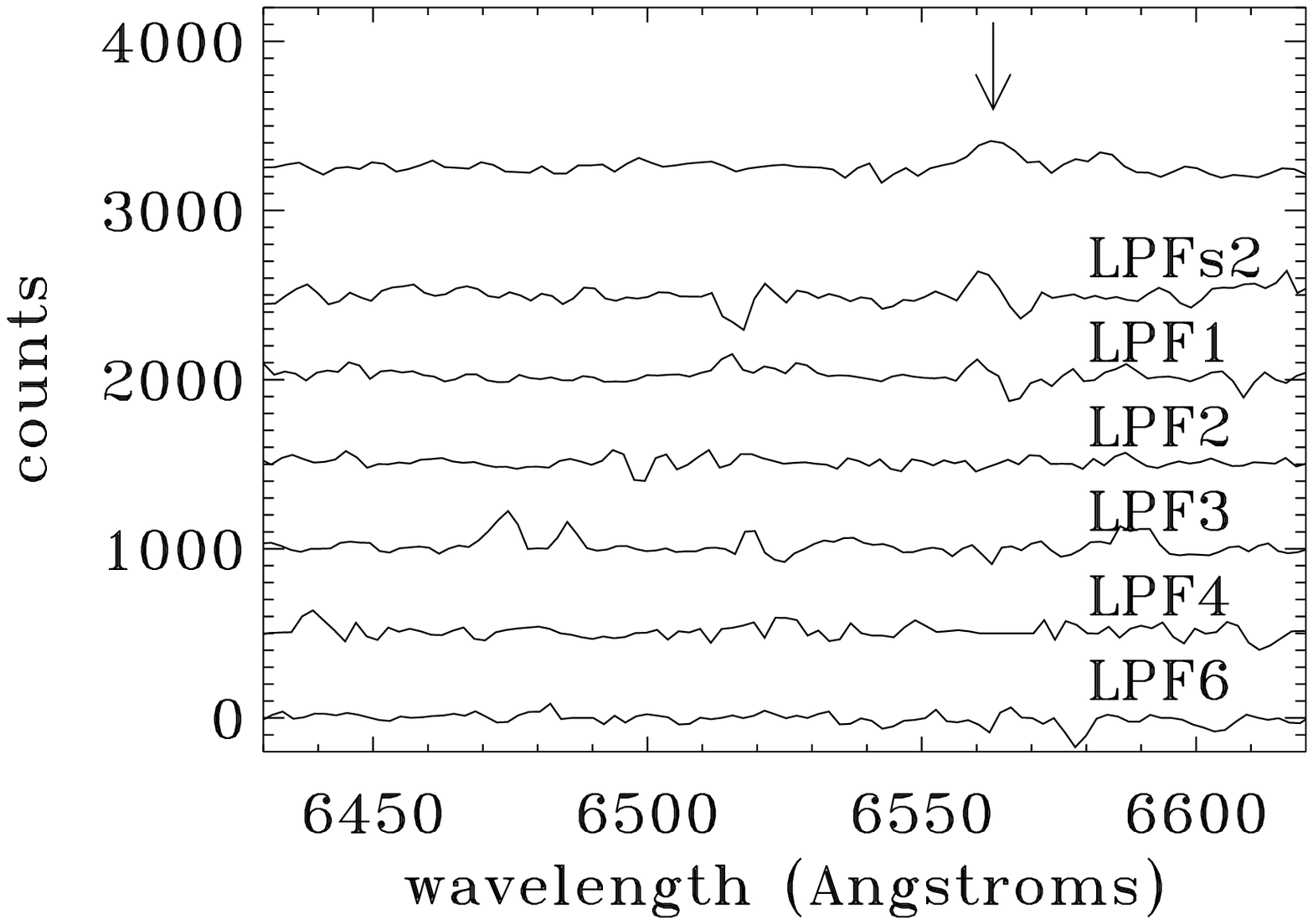]{The upper spectrum corresponds to another
anonymous galaxy, this time in Field 1, which shows emission at 
$\lambda$3727, $\lambda$5007 and H$\alpha$. The arrow indicates 
the position of H$\alpha$. Below it, from top to bottom, we show 
the spectra of 6 of our sources, all redshift-corrected as described 
in Fig.5. The levels of 
zero intensity are separated by 500 counts. None of these
sources shows any convincing evidence of emission at the expected
wavelength of H$\alpha$. The ``inverted P Cyg features'' in the
spectra of LPFs2 and LPF1 are artifacts produced by imperfect sky
subtraction in the presence of very strong sky emission lines.
We have verified by careful inspection of the combined CCD 
spectrograms that no real features are visible at those wavelengths. 
The spectra of the sources which are not shown here do not reach so 
large wavelengths, because
of their positions in their respective fields. This figure and the
previous one show that we cannot attribute the detected emission
lines to [O II] $\lambda$3727.
\label{fig6}}

\figcaption[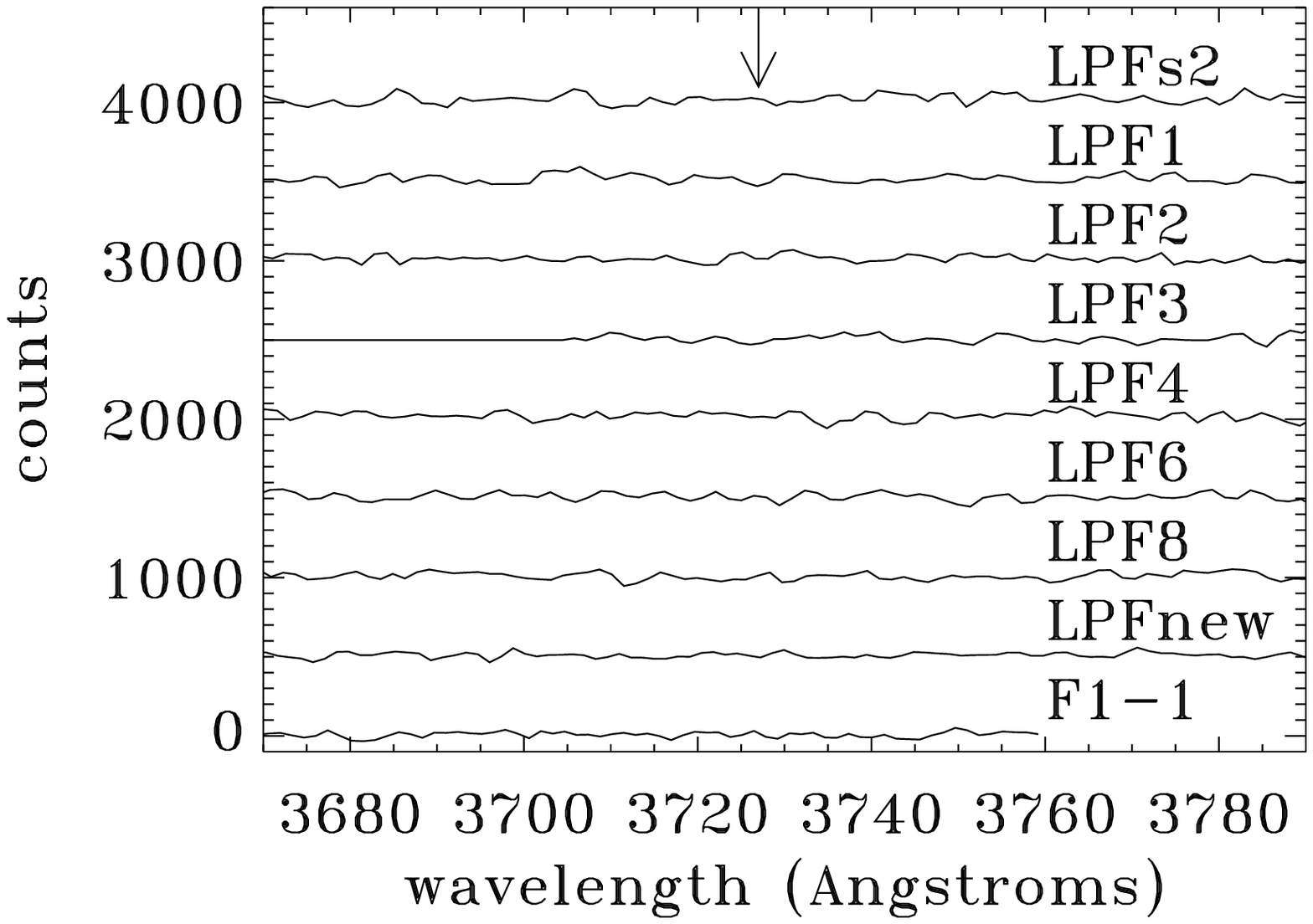]{Here all spectra have been redshift-corrected so 
that the strong emission line detected with the on-band filter falls 
at Mg II 2798 \AA. The levels of zero intensity are separated by 500 
counts. Nothing is visible at the expected position of 
[O II] $\lambda$3727, indicated with an arrow. \label{fig7}}

\figcaption[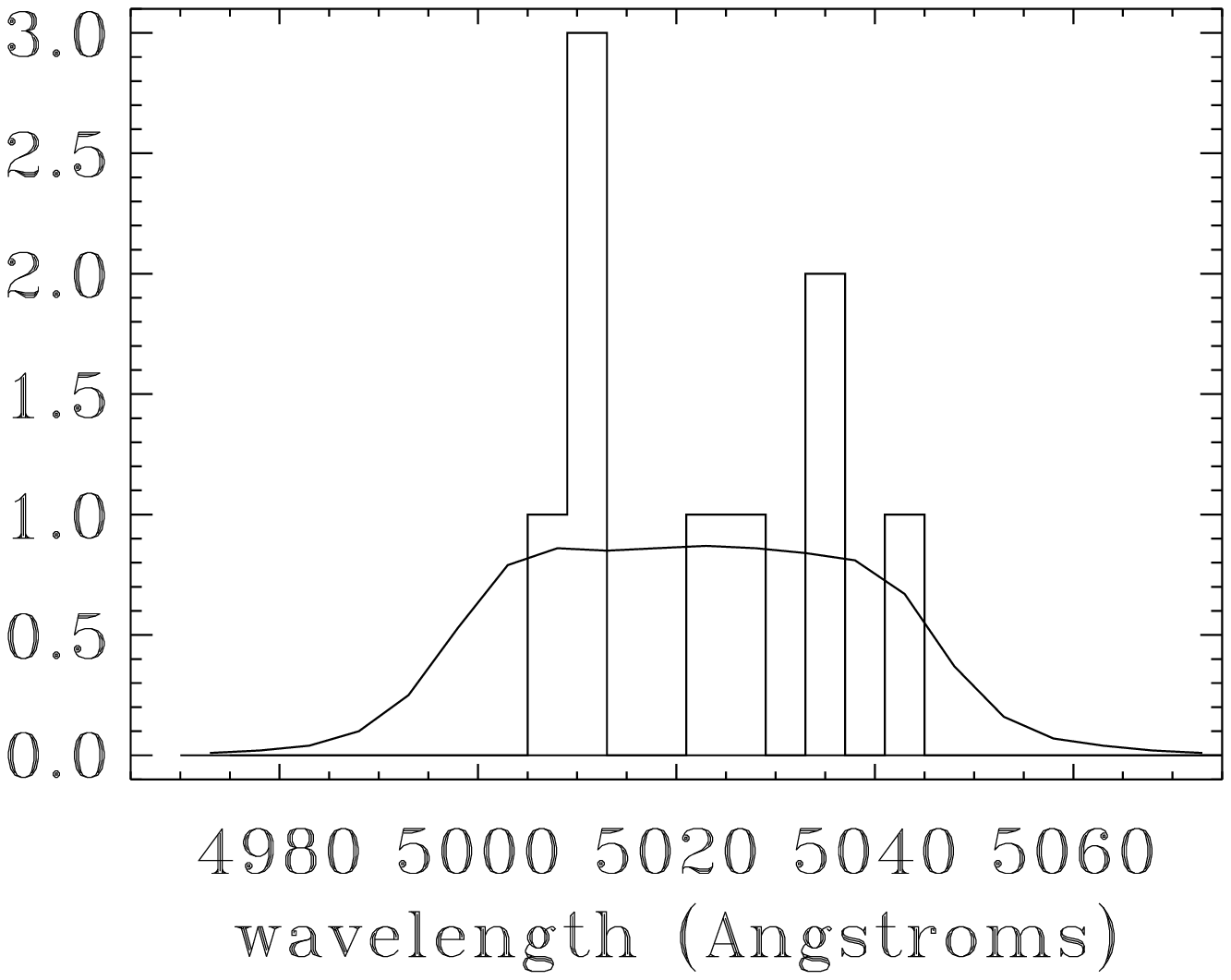]{The full line is the transmission curve of the
on-band filter used at La Palma. The histogram shows the wavelength
distribution of the emission lines. One object corresponds to 1 
unit along the y axis. Our sources are more or less uniformly 
distributed across the filter curve. \label{fig8}}

\figcaption[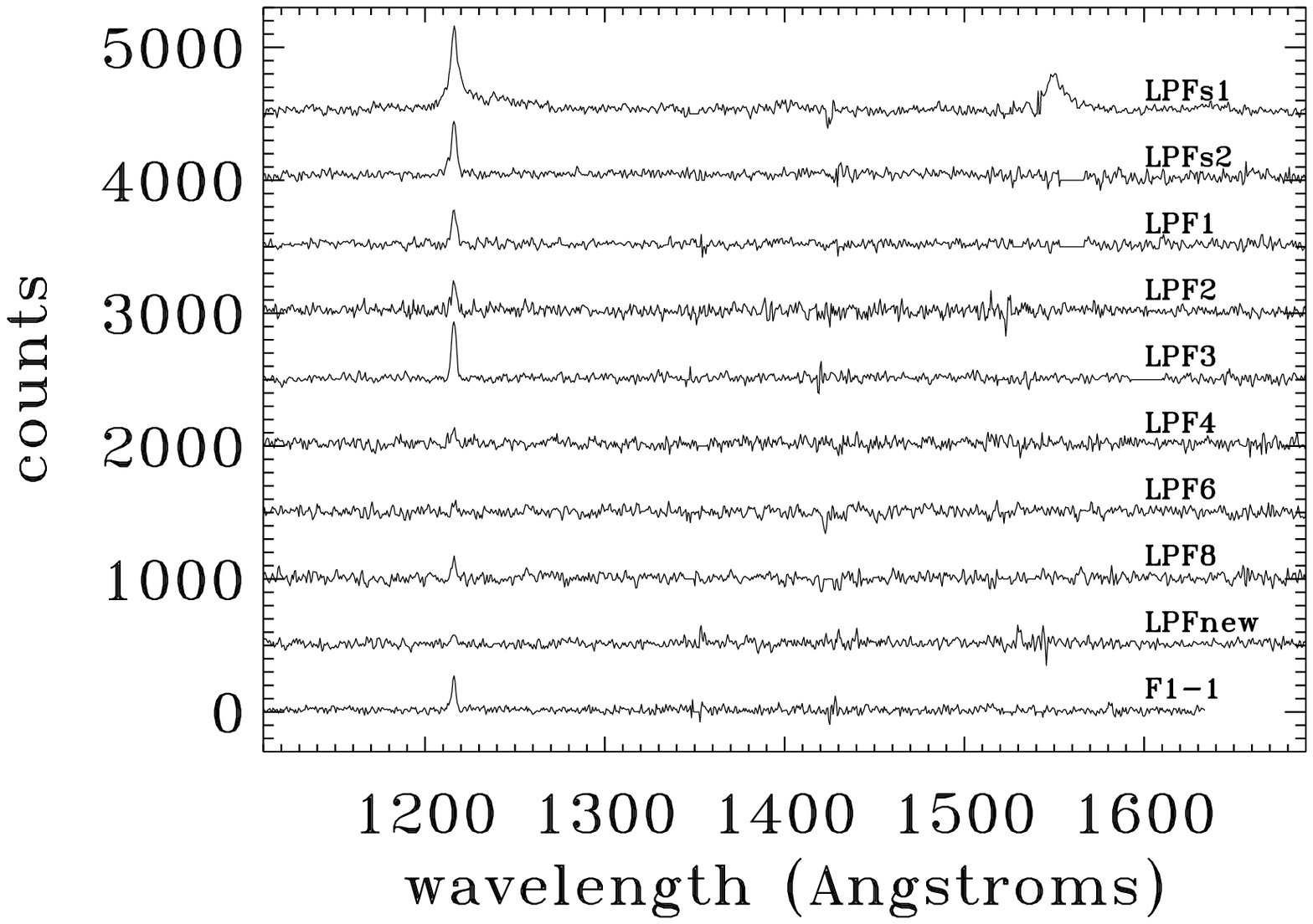]{Spectra of our sources in the rest frame of 
Ly$\alpha$. At the top is the QSO LPFs1. The levels of zero 
intensity are separated by 500 counts. Only the QSO shows C IV 
$\lambda$1550. At some wavelengths a slightly larger noise level 
is noticeable; this is due to imperfect sky subtraction in the
presence of very strong sky emission lines. In several cases where
the disturbance by the sky lines was too bad, we have replaced 
the spectrum by a straight line at zero intensity. \label{fig9}}

\figcaption[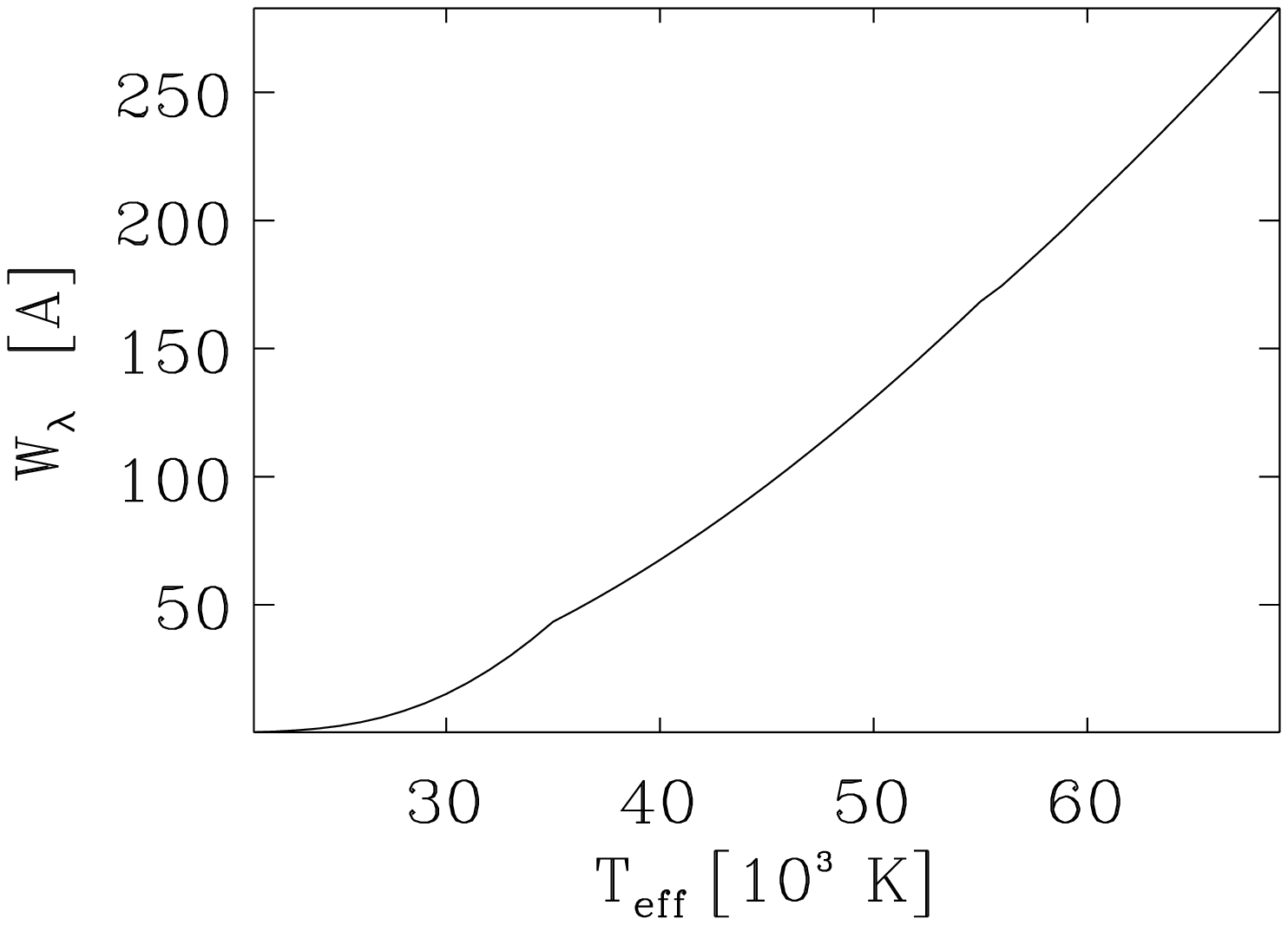]{Equivalent width of Ly$\alpha$ emission 
produced by a single star, as a function of the stellar 
T$_{\rm eff}$, calculated for $X_B$ = 0.3. \label{fig10}}

\figcaption[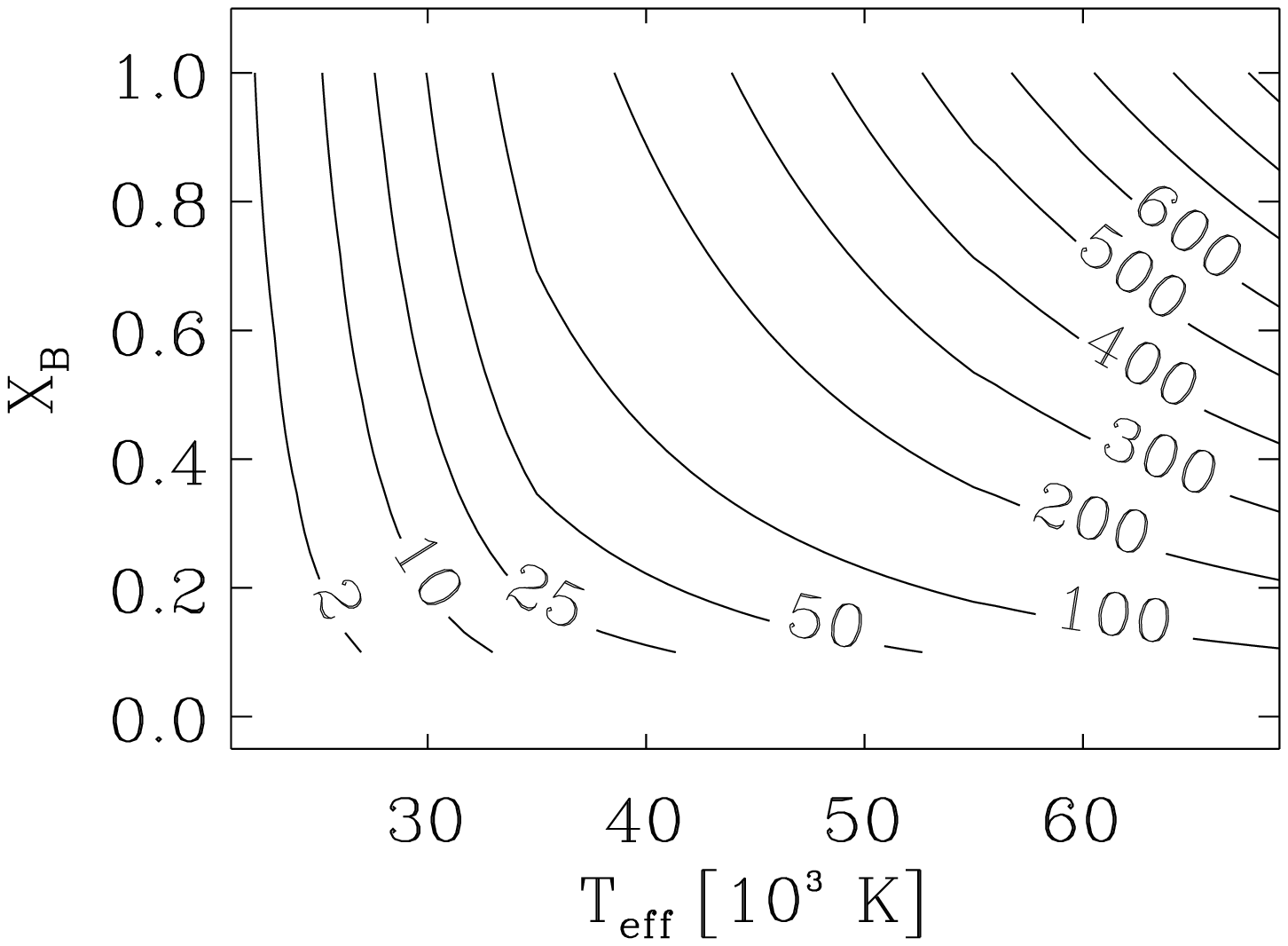]{ Equivalent width of Ly$\alpha$ emission 
produced by a single star, in the plane T$_{\rm eff}$ vs. $X_B$. 
The contour lines are labeled with the Ly$\alpha$ equivalent widths 
in \AA. \label{fig11}}

\figcaption[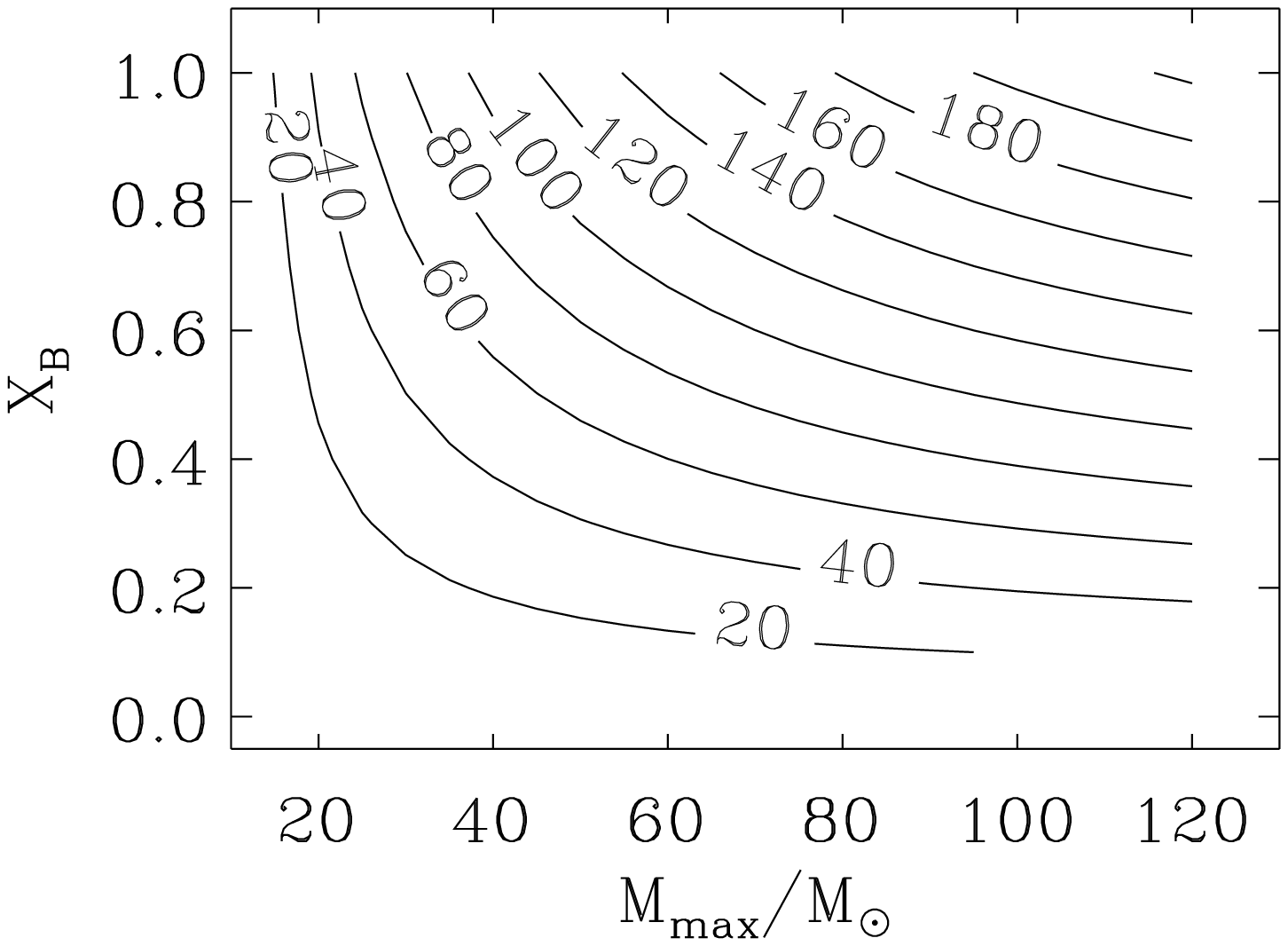]{Contour plots of Ly$\alpha$ equivalent widths
in \AA, as a function of $X_B$ and $M_{\rm max}$ (the maximum main 
sequence mass in the population), for the case of a starburst. 
See text. \label{fig12}}

\figcaption[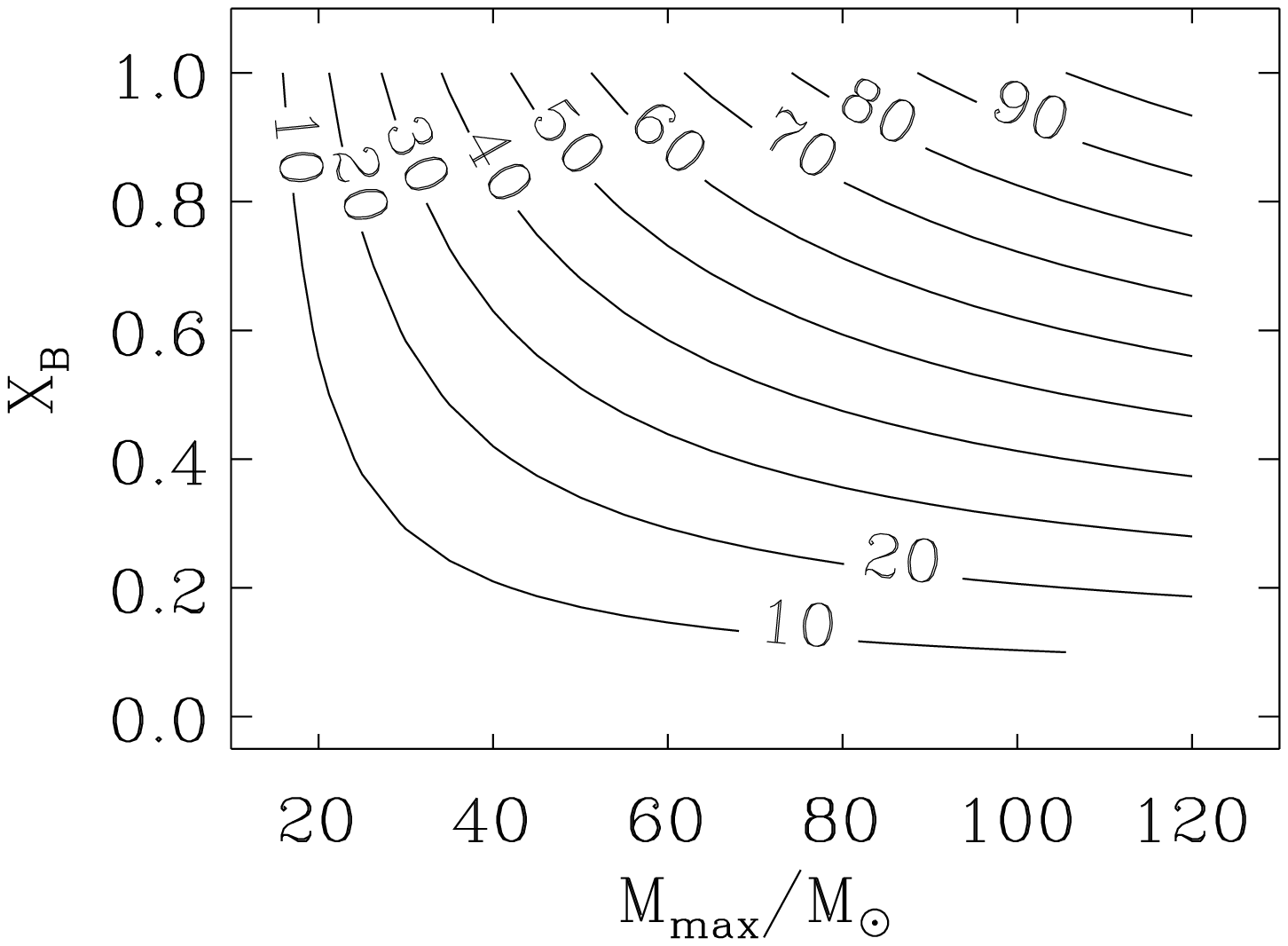]{Same as Fig. 12, for the case of continuous 
star formation. See text. \label{fig13}}

\figcaption[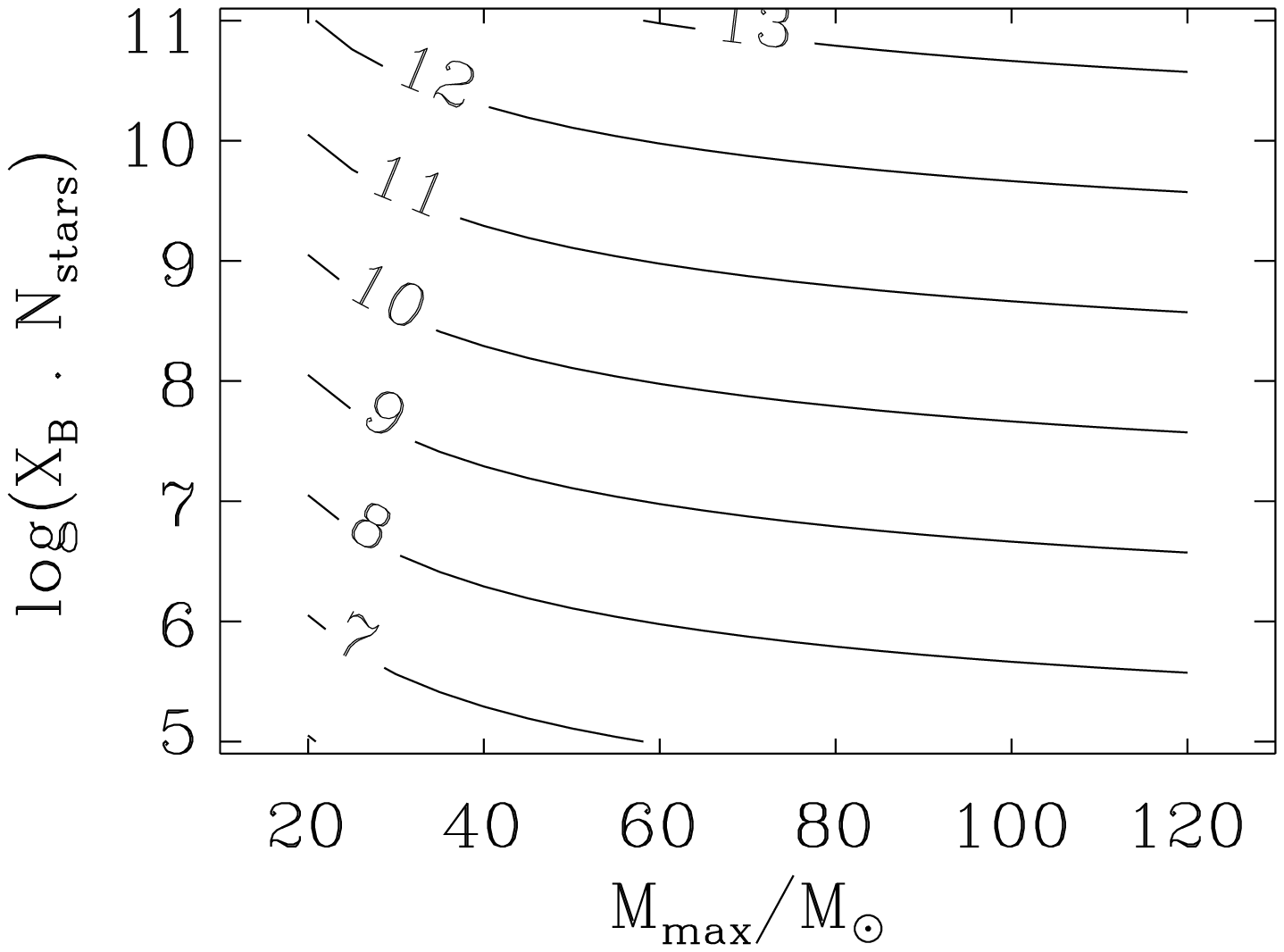]{Contour plots of the logarithms of Ly$\alpha$ 
luminosities, in units of L$_{\odot}$, for the case of a starburst.
Given a Ly$\alpha$ luminosity, it is possible to read, for different
values of $M_{\rm max}$, the product of $X_B$ times the total number
of stars. \label{fig14}}

\figcaption[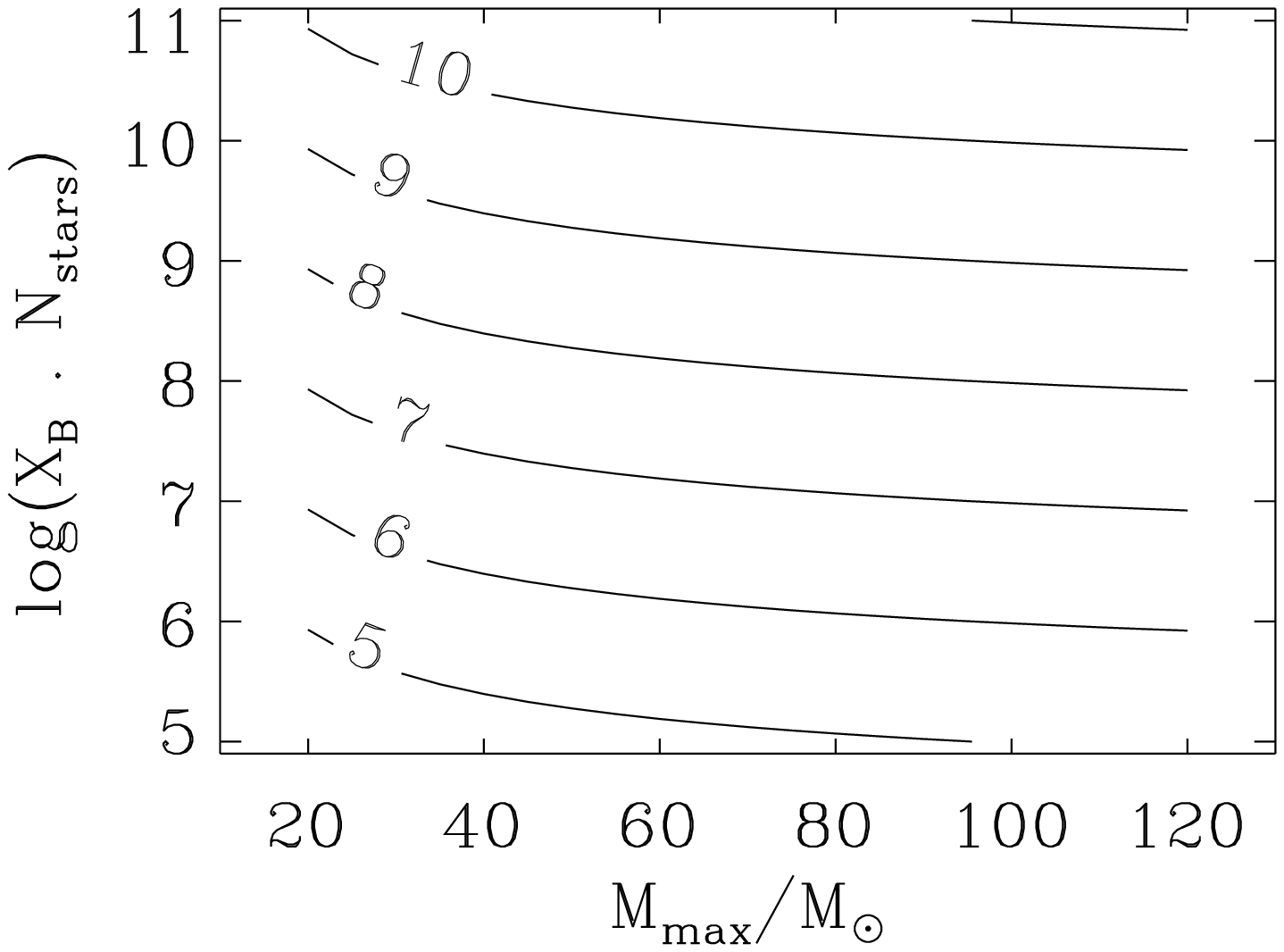]{Same as Fig. 14, for the case of continuous 
star formation. \label{fig15}}

\figcaption[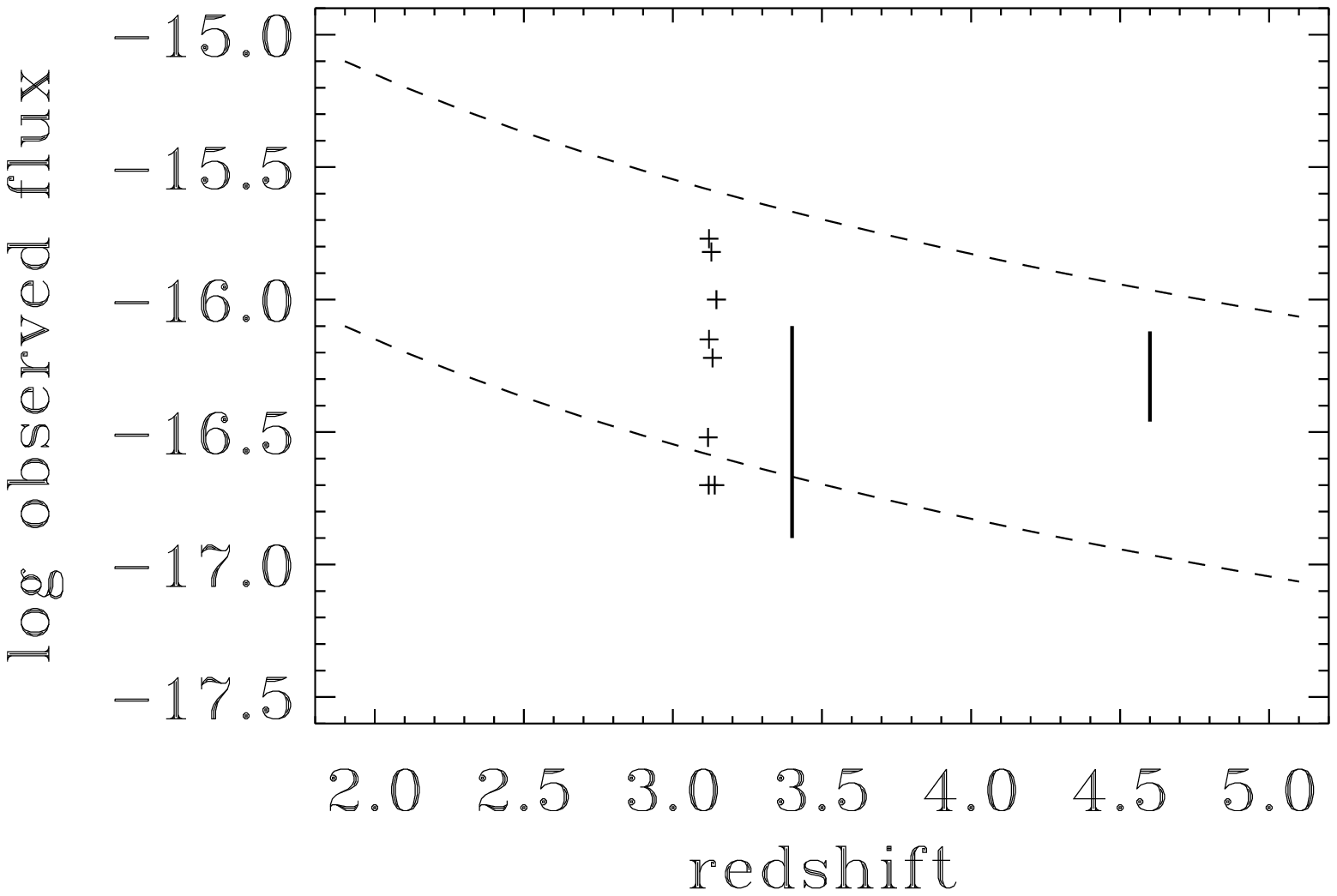]{The logarithms of observed Ly$\alpha$ fluxes
(in erg cm$^{-2}$ s$^{-1}$) as a function of redshift. The plus signs
are our sources, and the vertical bars represent the ranges of fluxes
observed by Hu (1998) at redshifts 3.4 and 4.6. The dashed lines 
represent the observed fluxes for Ly$\alpha$ luminosities of 
10$^{42}$ and 10$^{43}$ erg s$^{-1}$, if we assume
H$_0$=70 km s$^{-1}$ Mpc$^{-1}$ and q$_0$=0.5. \label{fig16}}

\figcaption[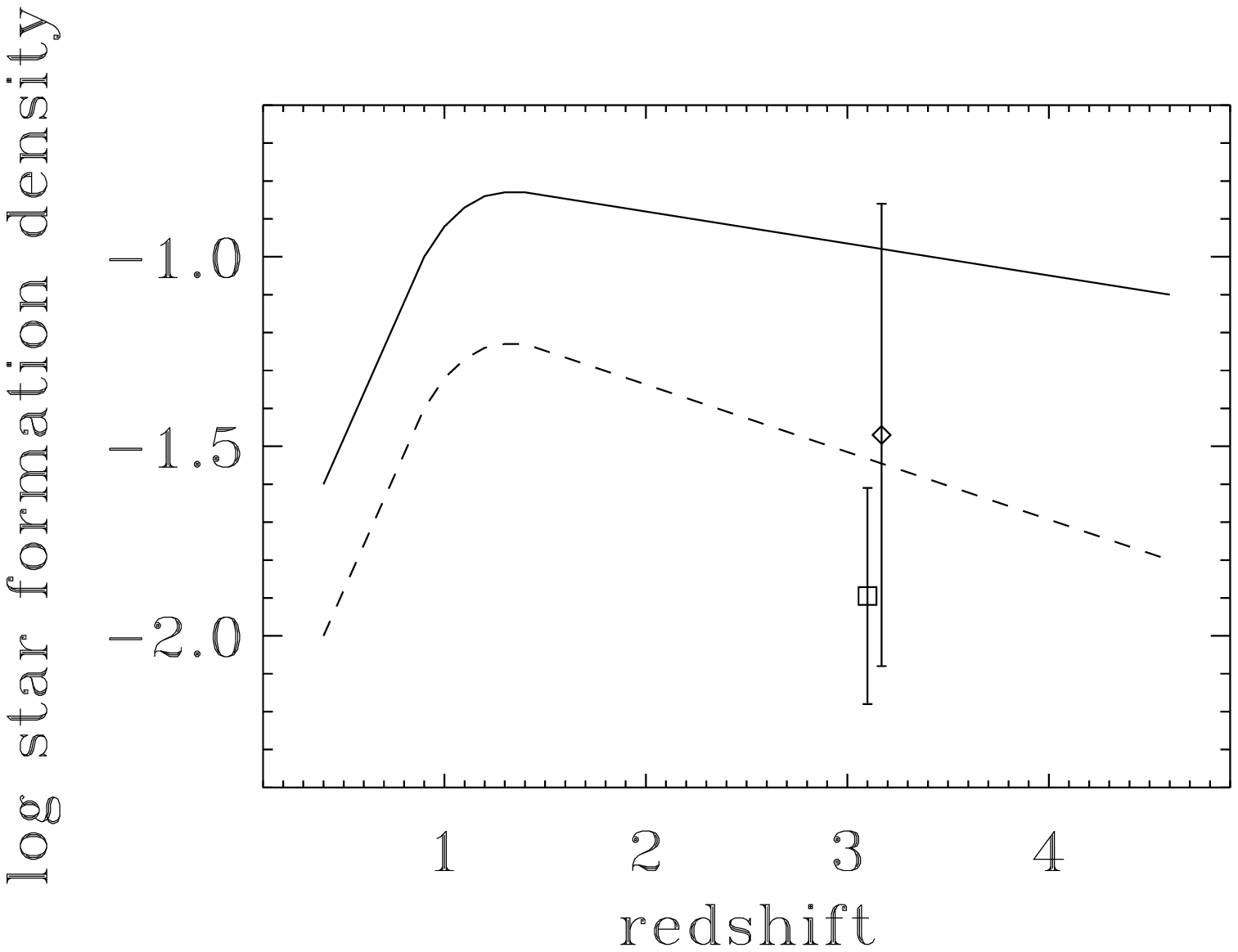]{Logarithms of star formation density in 
M$\odot$ yr$^{-1}$ Mpc$^{-3}$, as a
function of redshift. The lines show the results derived from galaxy 
surveys, collected by Steidel et al. (1999a): neglecting (dashed) and 
correcting for (solid) extinction by dust. These curves are derived
assuming continuous star formation. Our star formation densities at 
$z=3.13$ have been adapted, for consistency, to the same cosmological 
parameters used by Steidel et al. The square and diamond represent,
respectively, our results assuming continuous star formation and 
starbursts. Both symbols have been slightly displaced horizontally 
for legibility. \label{fig17}}

\newpage

\begin{figure}
\begin{center}
\figurenum{1}
\epsscale{1.0}
\plotone{fig1.ps}
\caption{Figure 1}
\end{center}
\end{figure}

\begin{figure}
\begin{center}
\figurenum{2 and 3}
\epsscale{1.0}
\plottwo{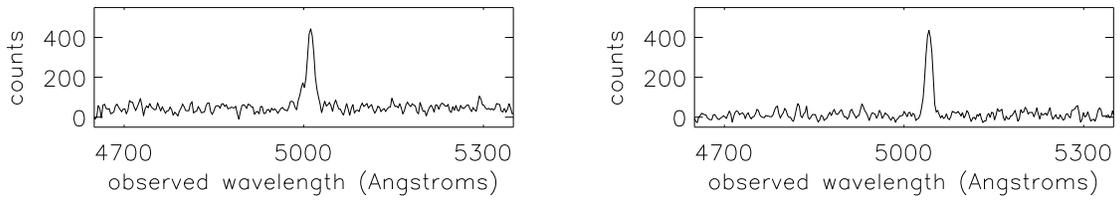}{fig3.ps}
\caption{Figures 2 and 3}
\end{center}
\end{figure}

\begin{figure}
\begin{center}
\figurenum{4}
\epsscale{1.0}
\plotone{fig4.ps}
\caption{Figure 4}
\end{center}
\end{figure}

\begin{figure}
\begin{center}
\figurenum{5}
\epsscale{1.0}
\plotone{fig5.ps}
\caption{Figure 5}
\end{center}
\end{figure}

\begin{figure}
\begin{center}
\figurenum{6 and 7}
\epsscale{1.0}
\plottwo{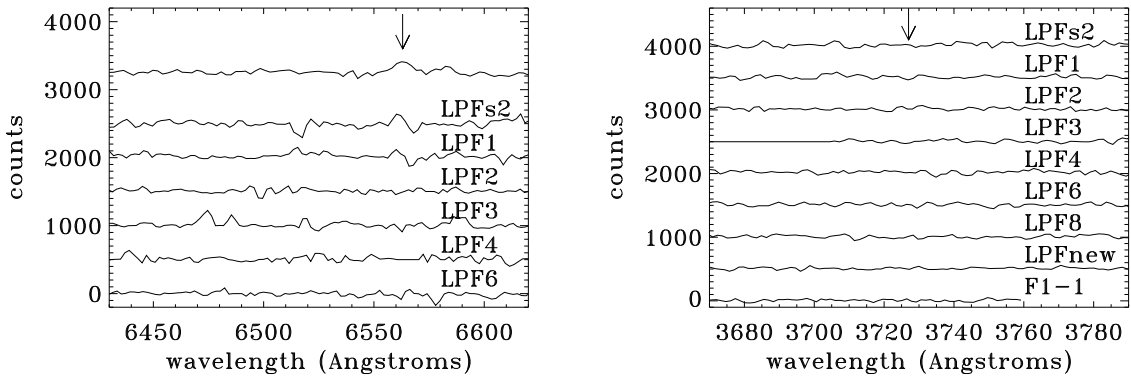}{fig7.ps}
\caption{Figures 6 and 7}
\end{center}
\end{figure}

\begin{figure}
\begin{center}
\figurenum{8}
\epsscale{1.0}
\plotone{fig8.ps}
\caption{Figure 8}
\end{center}
\end{figure}

\begin{figure}
\begin{center}
\figurenum{9}
\epsscale{1.0}
\plotone{fig9.ps}
\caption{Figure 9}
\end{center}
\end{figure}

\begin{figure}
\begin{center}
\figurenum{10 and 11}
\epsscale{1.0}
\plottwo{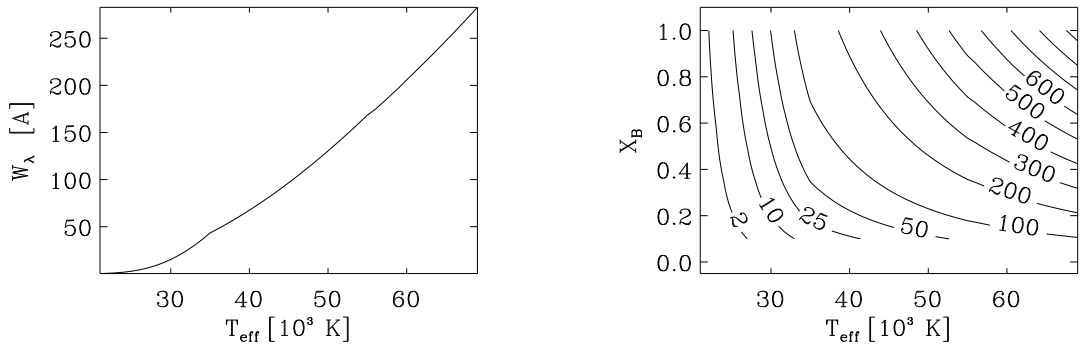}{fig11.ps}
\caption{Figures 10 and 11}
\end{center}
\end{figure}

\begin{figure}
\begin{center}
\figurenum{12 and 13}
\epsscale{1.0}
\plottwo{fig12.ps}{fig13.ps}
\caption{Figures 12 (starburst) and 13 (continuous star formation)}
\end{center}
\end{figure}

\begin{figure}
\begin{center}
\figurenum{14 and 15}
\epsscale{1.0}
\plottwo{fig14.ps}{fig15.ps}
\caption{Figures 14 (starburst) and 15 (continuous star formation)}
\end{center}
\end{figure}

\begin{figure}
\begin{center}
\figurenum{16 and 17}
\epsscale{1.0}
\plottwo{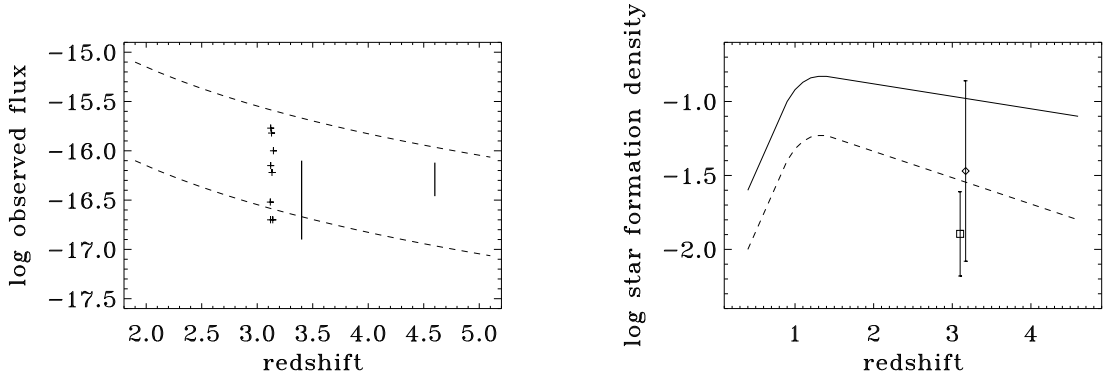}{fig17.ps}
\caption{Figures 16 and 17}
\end{center}
\end{figure}

\end{document}